\documentclass[twocolumn,doublespace,showpacs]{MIEEEtran}
\usepackage{amsmath,amssymb}

\newcommand{\comment}[1]{{}}

\newcommand{\smfrac}[2]{\mbox{$\frac{#1}{#2}$}}
\def\>{\rangle}
\def\<{\langle}
\newcommand{\eq}[1]{Eq.~(\ref{eq:#1})}
\def\e{\epsilon}
\def\SE{\mbox{{\sc se}}}
\def\E{\mbox{{\sc e}}}
\def\lpm{ \left(\rule{0pt}{2.1ex}\right. \!\! }
\def\rpm{ \!\! \left.\rule{0pt}{2.1ex}\right) }

\def\untrust{^{\mbox{\tiny ?}}\hspace*{-0.85 ex}\rho}
\def\good{\gamma}
\def\whpgood{\tilde{\gamma}}
\def\initial{\rho_{\rm 0}}

\newcommand{\CC}{{{\mathbb C}}}

\def\duzomniejsze{<\kern-.7mm<}
\def\duzowieksze{>\kern-.7mm>}

\def\textbf#1{{\bf #1}}

\def\bep{\begin{proposition}}
\def\eep{\end{proposition}}

\def\beq{\begin{equation}}
\def\eeq{\end{equation}}
\def\be{\begin{equation}}
\def\ee{\end{equation}}
\def\bea{\begin{eqnarray}}
\def\eea{\end{eqnarray}}
\def\beqa{\begin{eqnarray}}
\def\eeqa{\end{eqnarray}}
\def\ben{\begin{eqnarray}}
\def\een{\end{eqnarray}}

\newcommand{\bei}{\begin{itemize}}
\newcommand{\eei}{\end{itemize}}
\newcommand{\bee}{\begin{enumerate}}
\newcommand{\eee}{\end{enumerate}}

\def\hcal{{\cal H}}
\def\pcal{{\cal P}}
\def\acal{{\cal A}}

\def\mcal{{\cal M}}
\def\zcal{{\cal Z}}

\def\tr{{\rm Tr}}

\def\ra{\rightarrow}

\def\>{\rangle}
\def\<{\langle}

\def\ot{\otimes}

\def\ot{\otimes}  

\def\ep{\epsilon}

\def\ap_nr{$Sym(\hcal^{\ot n},|\theta\>^{\ot n-r})$}
\def\mix_ap_nr{$\pcal(Sym(\hcal^{\ot n},|\theta\>^{\ot n-r}))$}

\newtheorem{lemma}{Lemma}

\newtheorem{proposition}{Proposition}
\newtheorem{theorem}{Theorem}
\newtheorem{definition}{Definition}
\newtheorem{obs}{Observation}

\newtheorem{corollary}{Corollary}

\newtheorem{fact}{Fact}

\def\Sym{{\rm Sym}}
\def\Li{\langle L_{i} \rangle_{\sigma}}
\def\Romr{\varrho^{(\sigma)}_{2m,r}}
\def\Liemp{\langle L_{i} \rangle_{\rm emp}}
\def\Sgm{\langle\Sigma\rangle_{\sigma}}  
\def\SgmAemp{\langle\Sigma\rangle_{\rm emp}^{(m),{\rm dir}}}
\def\SgmAAemp{\langle\Sigma\rangle_{\rm emp}^{(m+n)}} 
\def\SgmBemp{\langle\Sigma\rangle_{\rm emp}^{(m),{\rm ind}}}

\def\vcal{{\cal V}}
\def\bec{\begin{corollary}}
\def\eec{\end{corollary}}

\def\bel{\begin{lemma}}
\def\eel{\end{lemma}}
\def\bet{\begin{theorem}}
\def\eet{\end{theorem}}

\def\almpower{\rho^{(\sigma)}_{n,r}}

\def\pmqkd{P/M-QKD\ }

\begin{document}

\title{Quantum key distribution based on private states:  
unconditional security over untrusted channels with zero quantum capacity}

\author{
Karol Horodecki, 
Micha\l{} Horodecki,  
Pawe\l{} Horodecki, 
Debbie Leung, and Jonathan Oppenheim
\thanks{Karol and Micha\l{} Horodecki were in the Department of
Mathematics, Physics and Computer Science, University of Gda\'nsk,
80--952 Gda\'nsk, Poland, Pawe\l{} Horodecki was in the Faculty of
Applied Physics and Mathematics, Technical University of Gda\'nsk,
80--952 Gda\'nsk, Poland, Debbie Leung was at the Institute for
Quantum Computing, University of Waterloo, Waterloo, Ontario, N2L1N8,
Canada, and Jonathan Oppenheim was at the Department of Applied
Mathematics and Theoretical Physics, University of Cambridge, U.K.} }

%%%%%%%%%%%%%%% note the following is not in use %%%%%%%%%%%%%%%%%%%%
\comment{
\author{Karol Horodecki$^{(1)}$, Micha\l{} Horodecki$^{(1)}$, Pawe\l{}
Horodecki$^{(2)}$, Debbie Leung$^{(3)}$, Jonathan Oppenheim$^{(4)}$}

\affiliation{$^{(1)}$Department of Mathematics, Physics and Computer
Science, University of Gda\'nsk, 80--952 Gda\'nsk, Poland}
\affiliation{$^{(2)}$Faculty of Applied Physics and Mathematics,
Technical University of Gda\'nsk, 80--952 Gda\'nsk, Poland}
\affiliation{$^{(3)}$Institute for Quantum Computing, University of 
Waterloo, Waterloo, Ontario, N2L1N8, Canada}
\affiliation{$^{(4)}$Department of Applied Mathematics and
Theoretical Physics, University of Cambridge, U.K.}
}
%%%%%%%%%%%%%%%%%%%%%%%%%%%%%%%%%%%%%%%%%%%%%%%%%%%%%%%%%%%%%%%%%%%%%%

\maketitle

\begin{abstract}

We prove unconditional security for a quantum key distribution (QKD)
protocol based on distilling pbits (twisted ebits) \cite{HHHO03} from
an arbitrary {\em untrusted} state that is claimed to contain
distillable key.
Our main result is that we can verify security using only public
communication -- via parameter estimation of the given untrusted
state.
The technique applies even to bound entangled states, thus extending
QKD to the regime where the available quantum channel has zero quantum
capacity.
We also show how to convert our purification-based QKD schemes to
prepare-measure schemes.
\end{abstract}

\parskip=1ex
\parindent=0ex

\section{Background, problem, and result} 

A large class of Quantum Key Distribution (QKD) protocols are based on
entanglement-purification-protocols (EPP).  
We use the shorthand EPP-QKD for these protocols.  
It is known that a secure key can be obtained by locally measuring two
systems prepared in some maximally entangled state (also known as EPR
pairs \cite{EPR} or ebits).  
The security and working principle of EPP-QKD are based on the ability
of two separated parties to estimate error rates of an {\em untrusted}
shared state relative to ebits and to subsequently distill ebits.
In \cite{HHHO03}, it was found that the most general quantum state
(known to the users) which provides a secure key (after measurement)
is not an ebit.  It is called a pbit or ``twisted ebits.''
These pbits can likewise be distilled or purified from known shared
states.
This paper is focused on the scenario when the users share {\em
untrusted} states, and how to devise QKD schemes under such
circumstances based on pbit-purification-protocols.  We call these 
protocols PPP-QKD.
The main goal is to devise an analoguous error estimation scheme
relative to pbits, using only public classical communication.
The scheme applies to some ``bound entangled'' initial states that are
nonetheless sufficiently close to pbits.  (A state is bound entangled
if no ebits can be distilled from many copies of it.)
Consequently, there are channels that cannot be used to send quantum
information (zero quantum capacity), but that can be used for QKD
(nonzero key capacity).  
Furthermore, in spirit similar to \cite{SP00}, we provide a recipe for
converting PPP-QKD to their associated prepare-measure schemes
(\pmqkd).
We will concentrate on the verification scheme of Lo, Chau, and
Ardehali \cite{LC99,LCA01} where bit and phase error rates are
estimated.  We instead estimate ``twisted'' bit and phase error rates.
Our proof uses classical random sampling theory, and the exponential
quantum de Finetti theorem \cite{Renner05}.

We first provide a pedagogical review on the essential concepts of QKD
in Sec.\ \ref{sec:review}.  Readers familiar with QKD can skip the
review.  We then discuss the current problem in Sec.\
\ref{sec:motivation} followed by a precise statement of our results in
Sec.\ \ref{sec:results} as well as related results in Sec.\
\ref{sec:related}.  The proof of security is contained in Sec.\
\ref{sec:details} with the essence of it being in Sec.\
\ref{sec:pstates}.
We follow this up by a discussion on how to convert our protocol to a
\pmqkd scheme in Sec.\ \ref{sec:pm} and give an example
of QKD using a binding-entanglement-channel in Sec.\ \ref{sec:example}
where the error rate is so high that quantum capacity vanishes.
An interesting observation that the users will not need to know about
what private state they share and how to exploit this fact are given
in Sec.\ \ref{sec:opu}.
We have built our protocols piece by piece, and a summary of the 
complete protocols is given in Sec.\ \ref{sec:sumprot}. 
We conclude with other remarks and a discussion of open problems in
Sec.\ \ref{sec:discussion}.  Proofs are detailed in the Appendix, and
the theorems are restated in the body of the paper.
  
\subsection{Review of quantum key distribution (QKD)} 
\label{sec:review}

In the quantum world, it is generally impossible to extract
information about a quantum state without disturbing it
\cite{Bennett94a}.
This principle enables {\em unconditionally secure key distribution}
that is impossible classically.  
Key distribution is the task of establishing a key between two
parties, Alice and Bob.
Informally, a key distribution protocol is secure if the probability
to establish a compromised key vanishes.  (In the above statement, we
have allowed the key length to vary and when it is zero, the protocol
``aborts''.  See also \cite{BHLMO04}.) If a protocol (given some
stated resources) is secure against the most powerful adversary (Eve)
limited only by laws of physics, its security is ``unconditional.''

In QKD, Alice and Bob can use a quantum channel (from Alice to Bob)
and classical channels (in both directions).  These can be noisy and
controlled by Eve.  In addition, Alice and Bob have local coins and in
some cases, quantum computers.  These can be noisy but they are not
controlled by Eve.  Finally, Alice and Bob share a small initial key.
Using the quantum universal composability result \cite{BM02}, most of
the imperfect resources can be made near-perfect while preserving
security -- the classical channels and local resources can be made
reliable and authenticated using the initial key and coding.
We make these simplifying assumptions from now on, and focus on 
imperfections in the quantum channel.   
(As a side remark, for arbitrary adversarial imperfections in the
quantum channel, no coding method can convert it to a perfect quantum
channel.  Fortunately, QKD requires less (see above) and this paper
revolves around the minimal requirement on the quantum channel.)

% Michal wants to add P/M, but it is more like a sentence ``The quantum 
% computer is not necessary in some case'' which is off the point is 
% this paragraph -- just to state the setting.  The above is a min fix

We first give the intuition behind the security offered by quantum
mechanics assuming a noiseless quantum channel.  
Alice and Bob pre-agree on a set of non-orthogonal quantum states,
each may be transmitted by Alice through the quantum channel with
some probabilities.  (This is the case in the earliest QKD scheme
called BB84 \cite{BB84}.)
Eve can intercept and compromise the quantum signals but they will be
disturbed.
Bob tells Alice when he receives the states, and they subsequently
detect disturbance using some of the states, and if they observe none,
they extract a key from the rest of the states; otherwise, they abort
the protocol.

QKD based on noisy channels is important for two reasons.  First,
natural noise is inevitable and can be used by an eavesdropper as a
disguise.  Second, it is desirable to be able to generate a key
despite some malicious attack.
Initial work \cite{Bruss98,DEJMPS96} was done based on error
estimation, privacy amplification \cite{BBCM95}, and error correction.
Mayers first gave an unconditional security proof for QKD
\cite{Mayers96}, showing that BB84 can provide a key up to $\approx 8
\%$ observed error.

Later on, Lo and Chau \cite{LC99} reported a security proof for a
different QKD scheme based on E91 \cite{E91} -- Alice and Bob first share
some noisy, untrusted state $\untrust$.  
(We tag the state with a ``question mark'' to emphasize that the users 
cannot ascertain its identity.) 
It is supposed to be $n$ copies of $|\Phi_d\> := \smfrac{1}{\sqrt{d}}
\sum_{i=1}^d |i\>_A |i\>_B$ where $\{|i\>\}$ is a computational basis
for the local systems $A$ and $B$ possessed by Alice and Bob
respectively.
$|\Phi_d\>$ is called a ``maximally entangled state'' or MES for short.
When $d=2$ it is also called an EPR pair or ``ebit''. 
$\untrust$ arises from Alice preparing $n$ {\em local} copies of
$|\Phi_d\>$ and transmitting Bob's shares through an untrusted channel
of $d$ dimensions.  Eve can attack on all $n$ systems jointly.
For now, we focus on the $d=2$ case (just like \cite{LC99}).  
After Bob receives the state, Alice and Bob extract a smaller number
$m$ of nearly perfect ebits, from which a key is obtained by measuring
in the local computational basis.  It is possible for $m=0$, when QKD
is aborted.
The Lo-Chau proof is simple -- however the noise arises, just detect
and remove it, and doing so only involves standard techniques in
entanglement purification protocols (EPP) or distillation
\cite{BDSW96}.

The disadvantage of the Lo-Chau proof is that, its associated scheme
requires quantum storage and coherent manipulation of quantum data,
neither of which is required in BB84. 
Shor and Preskill \cite{SP00} provided a recipe to relate BB84 to the
E91-Lo-Chau scheme, such that the security of the former is implied by
that of the latter.  Furthermore, \cite{SP00} generalizes to many
other variants of BB84 (collectively called ``prepare-measure'' scheme
\pmqkd) so that their security can be proved via that of a related
purification-based QKD scheme.

\subsection{Step-by-step QKD and motivation of current problem}
\label{sec:motivation}

We discuss useful general concepts by interpreting the Lo-Chau
scheme~\cite{LC99} as follows.  
Alice and Bob preagree on a set of parameters $e$ for states, and let
sets of states sharing the same parameters be labeled as $S_e$.  (It
will be clear later how they should be chosen.)
The protocol is a 4-step process for Alice and Bob: 
\\
(1) Distribute an untrusted bipartite state $\untrust$ using the
    untrusted resources. \\
(2) Perform tests (via public discussion) on $\untrust$ such that if
    $\untrust \in S_e$, the test will output $e$ with high
    probability.  \\
    They only need to know which $e$ (or $S_e$) but not which
    $\untrust$, and the remaining procedure depends only on $e$ and
    applies to all states in $S_e$.  For example, in the Lo-Chau
    scheme, $e$ consists of two error rates (bit and phase), $S_e$ is
    the set of states arising from inflicting errors of rates $e$ to
    the Bell states (see Def.\ \ref{def:ege} and \eq{okerror} towards 
    the end of Sec.\ \ref{sec:tolerableattacks} for a
    precise definition). \\
(3) Based on the parameter $e$, apply an appropriate EPP to $\untrust$
    and output a state $\whpgood$. \\
    This procedure, if applied to any state in $S_e$, will return a
    state $\tilde{\gamma}$ which is a good approximation of a known and
    trusted state $\good$ (e.g.\ ebits in E91/Lo-Chau).  \\
% 
% NB the approx is the near-perfect ebits (or pbits later) 
% 
(4) Generate a key by measuring $\whpgood$ locally.\\
    The key can have varying size (depends on $e$), and zero
    key-length means ``abort QKD.''

We will refer to these $4$ steps (and their variations) repeatedly
throughout the paper.

To generalize the Lo-Chau scheme, we examine the requirements for each
of these steps (in reverse order).

We start with {\bf step (4): Simply suppose Alice and Bob share a
known and trusted state $\good$.  What $\good$ (other than
$|\Phi_d\>$) will generate a secure key?}  Reference \cite{HHHO03}
characterizes all such $\good$ (up to local unitaries on $A$ and $B$):
% 
% wcl: reusing d ...  
% 
\bea
        \gamma_d^U & = & U (\Phi_{dAB} \ot \rho_{A'B'}) U^\dagger 
\label{eq:gamma}
\\
        U & = & \sum_{ij} |ij\>\<ij|_{AB} \ot U_{ijA'B'}
\label{eq:twist}
\eea
where $\Phi_d = |\Phi_d\>\<\Phi_d|$ is the MES of local dimension $d$,
the subscripts $AA'$ and $BB'$ denote systems held by Alice and Bob,
$U_{ij}$ are unitary so that $U$ is also unitary, and $\rho_{A'B'}$ is
any state (pure or mixed) of some arbitrary dimension $d'$.  (Note
that dim$(\gamma_d^U) = d^2 d'$ and the key generated has size $\log
d$.)
$U$ in \eq{twist} is called the twisting operator, and any
$\gamma_d^U$ given by \eq{gamma} is called a pdit (or private state,
or twisted state or gamma state).
In some sense, \eq{gamma} and \eq{twist} characterize all the noise on
an MES that is {\em harmless} for the purpose of generating a key.
We state this property for the twisting operator $U$ more precisely.
\begin{obs} (See \cite{HHHO03,huge-key}) 
\label{th:ccq}
Let $U$ be any twisting operator, and consider any two states
$\rho_{AA'BB'}$, $\sigma_{AA'BB'}$ related by
$\sigma_{AA'BB'}=U\rho_{AA'BB'}U^{\dagger}$.  Let both be purified by
$E$.  (See definition in Sec.\ \ref{sec:lochau} right before
\eq{lochau-pure-1}.)
Then, if Alice and Bob measure $A$ and $B$ in the computational basis,
the reduced states on $ABE$, $\tilde{\rho}_{ABE}$ and
$\tilde{\sigma}_{ABE}$, are the same.
\end{obs}
Such postmeasurement states are called ccq states, for Alice and Bob 
hold classical systems, while Eve's state remains unmeasured and quantum.

Now consider {\bf step (3): Which states can be converted into a good
approximation of a private state} ($\good$)?  We call such states
``key distillable'' (even though they are not of tensor power form). 
The conversion procedure has to work for all states in $S_e$.  A
complete characterization is unlikely to be tractable and we only have
examples.
The canonical example in \cite{LC99} is the set of states with
sufficiently low error rates relative to perfect ebits.  We will call
these ``$\epsilon$-good-ebits''.  These are not necessarily tensor
power or product states (see Def.\ \ref{def:ege} and \eq{okerror}).
Here, EPP works for all $\epsilon$-good-ebits independent of which one
is the initial state.
Another example are tensor power states $\sigma^{\otimes n}$.
%
% DL: let's not mentioned known or nothing -- a distraction and both 
% are equivalent anyways. 
% 
% (with two-way communication and quantum memory, knowledge of $\sigma$
%can be obtained and need not be provided upfront).
% 
In this case, one says that a protocol achieves a ``key rate'' $r$ if
it converts $\sigma^{\otimes n}$ to $\approx \gamma_{2^{nr}}^U$ for
some $U$ given by \eq{twist}, allowing $n$ to be asymptotically large.
For example, protocols and lower bounds for $r$ are found in
\cite{DW03} for general $\sigma$.

We mention some surprising facts about private states and
key-distillable states.
All perfect pdits ($\good$) contain some distillable entanglement.
However, there are families of pdits with vanishing amount of
distillable entanglement but can be used to provide a key with
constant rate.  Also, there are states close to pdits and have
distillable key (lower bound from \cite{DW03}) but have no distillable
entanglement (upper bound from showing the positivity of the partial
tranpose (PPT)~\cite{Peres96,9801069}).
%(Here,more and more copies of $\sigma$ are used to approximate larger and
%larger private states, each is distillable.)

Switching from trusted states to untrusted states, we now move on to
the main concern of this paper: {\bf In step (2),
what sets $S_e$ contain key distillable states and admit parameter
estimation?  What are the corresponding tests for finding if $\untrust
\in S_e$?}
In the Lo-Chau proof, $\good$ are ebits and $S_e$ can be chosen to be
$\epsilon$-good-ebits (these are states with bounded error relative to 
ebits, see Def.\ \ref{def:ege} and \eq{okerror}).
In the most general case, $\good$ is a pdit and a natural question is,
can all key-distillable $S_e$ be tested?

We believe that the above question is hard, by considering all
possible ``$\epsilon$-good-pbits'' -- states obtained from applying
any twisting operation to $\epsilon$-good-ebits, where the twisting
can act jointly on the entire system.
Without further restriction on the joint twisting operation, it is
unclear how to perform parameter estimation on the joint state.

One particularly useful class of $\epsilon$-good-pbits are those
obtained from applying tensor-power twisting $U^{\ot n}$ on
$\epsilon$-good-ebits.  We will call these states
$\otimes$-twisted-$\epsilon$-good-pbits (note that just like
$\epsilon$-good-ebits, $\otimes$-twisted-$\epsilon$-good-pbits need not
be tensor power states).  (See also Def.\ \ref{def:egp}.)

Prior to this work, Ref.~\cite{HLLO05} showed how to perform parameter
estimation for $S_e$ containing $\otimes$-twisted-$\epsilon$-good-pbits
that have some distillable entanglement.
There, the important distinction from the Lo-Chau scheme is that, in
Ref.~\cite{HLLO05}, entanglement is only distilled for parameter
estimation but not for the subsequent key generation.  In particular,
the entanglement distilled in the scheme of Ref.~\cite{HLLO05} can be
in negligible quantity compared to the key size.
But \cite{HLLO05} leaves many questions unanswered, in particular,
whether $S_e$ can contain bound entangled but key distillable states,
and whether distilling entanglement (albeit a little) is necessary.
Also, the test in \cite{HLLO05} prevents easy conversion to a simpler
class of schemes called prepare-measure schemes (P/M-QKD, see below).

In this paper, we will show by an explicit protocol that parameter
estimation is possible {\em for all} $S_e$ containing states which can
be converted into $\otimes$-twisted-$\epsilon$-good-pbits by LOCC
operations (involving only local operations and public classical
communications).  Furthermore, this new estimation procedure does not
involve distillation so that it applies to bound entangled states; it
only involves product observables (see Def.\ \ref{def:prodobs}),
allowing easy conversion to P/M-QKD, as we will see later.

\subsection{Our adversarial setting} 
\label{sec:adv}

Throughout the paper, we are concerned with unconditional security of
QKD, in which nothing is assumed about the actual channel used or
about the actual state shared $\untrust$.  There are three separate 
notions that we want to mention explicitly.  

$\bullet$ Alice and Bob use an underlying quantum resource ``{\sc a}''
(a channel or quantum state) in order to execute QKD.  They have some
knowledge about this resource, for example, the natural channel loss
due to their distance can be theoretically calculated.

$\bullet$ During any specific execution, this resource is subject to
further unknown attack to produce the actual channel or state
$\untrust$ ``{\sc b}''.  This will remain unknown to Alice and Bob
throughout.

$\bullet$ What is known to Alice and Bob in an execution is a set of
observed error rates ``{\sc c}''.

The {\em insecurity} of QKD can be quantified by the probability that
the state has been compromised more than the observed error rates have
suggested.  Security is a consequence of the test procedure to obtain
``{\sc c}'', and is independent of any of the above.

It is a combination of the QKD protocol and the observed error ``{\sc
c}'' that determines the actual key rate, and this depends on the
actual channel or state $\untrust$ ``{\sc b}'', which in turns cannot
be better than the underlying resource ``{\sc a}''.
This is why analysis of a QKD protocol often refer to the underlying
resource ``{\sc a}'' -- the protocol, starting with resource ``{\sc
a}'' and subject to further unknown attack, will result in some
potentially worse observed error rates that may still give the
greenlight to establish a secure key.  $S_e$ can be used to describe
both concepts ``{\sc a}'' and ``{\sc c}''.

To repeat, given resource ``{\sc a}'' that is too noisy, QKD gives
zero key-rate whether there is eavesdropping or not.  On the other
hand, no matter how good ``{\sc a}'' is, too much eavesdropping should
also give a zero key-rate.  So, QKD is only interesting given good
enough underlying quantum resource ``{\sc a}'' together with a scheme
to ensure security.  The goal of this paper can be understood as
characterizing what underlying resources ``{\sc a}'' are good enough
under our scheme.

\subsection{Statement of results} 
\label{sec:results}

In this paper, we report a new test procedure in step (2) for any
$S_e$ containing $\otimes$-twisted-$\epsilon$-good-pbits.
Since Alice and Bob can use LOCC in QKD, our procedure also applies to
any $S_e$ containing states which can be converted into
$\otimes$-twisted-$\epsilon$-good-pbits by LOCC.
In particular, these include ({\sc 1})
$\otimes$-twisted-$\epsilon$-good-pbits themselves, and ({\sc 2}) tensor
power key distillable states.
This new method does {\em not} require distilling entanglement and it
applies independent of whether $S_e$ has distillable entanglement or
is bound entangled.

The protocol in this paper is similar to that in \cite{HLLO05}, and is
also a ``twist'' from the original Lo-Chau scheme.  
In the critical step of phase error estimation, we test for ``twisted
phase errors'' (phase errors in the basis defined by the twisting
operation) just as in \cite{HLLO05}.
In \cite{HLLO05}, the test is based on entanglement distillation and
teleportation.  Here, our new procedure uses a more recently found
finite quantum de Finetti theorem with exponential convergence
\cite{Renner05} and requires only local resources, measurement of
product observables (Def.\ \ref{def:prodobs}), and classical
communication.  This has significant consequences:

{\em (1) There are quantum channels that have zero quantum capacity but
nonzero key capacity.}  
Each set of states $S_e$ captures what Alice and Bob expect $\untrust$
to be.  It summarizes deviations from perfect pbits including channel
noise and noise inflicted by eavesdropping.
For example, Alice and Bob can have prior knowledge of the presumably
available quantum channel (resource (a) in previous subsection), which
is susceptible to further attack by an eavesdropper.  Our work extends
QKD to the regime when this presumably available quantum channel
cannot transmit quantum data, and only allows sharing of a bound
entangled key distillable state at best (without an eavesdropper).
With the unknown eavesdropping attack, the final key rate depends on
the combined noise level, and can potentially be positive.  (In the
static case, there are states that are untrusted and presumably bound
entangled that can still give a secure key.)

% FIX: IS SOME OF THE P/M STUFF REDUNDANT? 

{\em (2) Prepare-measure scheme based on private states}.  Remarkably,
in the noiseless case, E91 is mathematically related to many
``prepare-measure'' QKD schemes (P/M-QKD) including BB84.  P/M-QKD
only requires quantum states to be prepared and be sent by Alice, and
be measured by Bob without being stored, thus, minimal coherent
quantum manipulations.
P/M-QKD has much practical advantage over distillation or purification
based schemes, but the latter often have simple unconditional security
proofs.  Shor and Preskill \cite{SP00} illustrated mathematical
connections between the two types of schemes even in the noisy case
for some EPP.  Starting from the Lo-Chau security proof, they
rederived one for BB84 similar to Mayer's.  Reference \cite{GL01}
generalized the connection to more general EPP.
Likewise, our new test procedure allows the purification-based scheme
to be transformed to a P/M-QKD.  This can be useful in implementation.

We note a side result that may be of independent interest -- that
average values of an observable in a bulk system can be estimated in a
sublinear sample even when the observable cannot be directly measured.
This will be discussed more in Section \ref{sec:discussion}.

\subsection{Related work} 
\label{sec:related}

As already noted, this paper is a follow-up of \cite{HLLO05} on
parameter estimation of untrusted states relative to pbits.  The
scheme in \cite{HLLO05} requires a small amount of distillable
entanglement -- it does not apply to bound entangled states and thus
cannot be used on states generated by a channel with zero quantum
capacity.  

An earlier version of the current result (unpublished) used an exact
but polynomial quantum de Finetti theorem \cite{RK04b} from which we
obtained a much lower key rate.  The new exponential quantum de
Finetti theorem (exp-QDFT) in \cite{Renner05} provides much better
bounds and properties.

There are two intuitive solutions to the current problem of parameter
estimation.  The first is a state-tomographic estimation, which was
suggested in \cite{HHHO03}, but the accuracy and security was not
analyzed.  It is interesting to note here that exp-QDFT provides
exactly the tool for doing so.  Whatever $\untrust$ is, Alice and Bob
can simply choose half of the systems (or any linear amount) at
random, and the chosen state $\untrust'$ is exponentially close to a
mixture of ``almost-power-state.''  $S_e$ can be chosen to be tensor
powers of key distillable states and the test for $\untrust' \in S_e$
simply involves state tomography using only measurement of product
observables and classical communication.  (During the final
preparation of the manuscript, we heard of some work in progress using
this approach \cite{JC}.)
This paper follows another intuitive approach -- error estimation in
the twisted basis, via a decomposition of the twisted observable into
product observables (see Def.\ \ref{def:prodobs}).  Intriguingly, a
natural choice of the set of product observables is also
tomographically complete.  However, discarding is not necessary here.
The main challenge is a rigorous security proof, along with a careful
analysis of how various parameters are related.  We have used many
different elements (including the exp-QDFT) in \cite{Renner05}, along
with earlier techniques such as quantum-classical-reduction and
various random sampling techniques, \cite{LC99,LCA01}, and also ideas
from \cite{HLLO05}.

In this paper, we have also emphasized various useful concepts, such
as ``harmless errors'' and the structural constituents of QKD.
Examples of harmless errors (most generally defined by the private
states) was observed in earlier works by Aschauer and Briegel
\cite{AB02} and was used in \cite{RGK05,PL05,RSS06} to improve the key
rate.
Various useful structural descriptions of QKD, revolving more around
P/M-QKD, have also been proposed before \cite{LCA01,CRE04,Renner05}.

After the initial presentation of this result \cite{QIP06}, and during
the preparation of the current manuscript, Renes and Smith \cite{RS06}
reported the following related result.  The P/M-QKD scheme
\cite{RGK05,KGR05} that uses local noise inflicted by Alice to
increase the key rate has an interpretation as a QKD scheme based on
distributing and distilling a particular private/twisted state.  Thus
they arrived at an (existing) example of P/M-QKD based on private
states (but the state has to be (ebit) distillable since the noise is
local).  This is complementary to our current result (item (2)) that
aims at a general recipe to convert distillation-based schemes to
P/M-QKD.

Finally, \cite{bigkey-shortpaper} contains a summary of this paper
without the technical details.

\section{Details of our result} 
\label{sec:details}

Recall that in the current formulation of QKD, the goal is to
accurately test whether the shared bipartite state $\untrust$ is in
some set $S_e$ or not, and if so, apply a transformation that will
bring any state in that $S_e$ to a state close to a private state
$\good$.
The test and transformation use only LOCC.   
Note that $\untrust$ is determined by eavesdropping and the channel
properties, while $S_e$ and the test is part of the design of the QKD
scheme.
We will describe and prove the security of a QKD scheme with $S_e$
containing $\otimes$-twisted-$\epsilon$-good-pbits (see Def.\
\ref{def:egp}).

As described before, our procedure also applies to any $S_e$
containing states which can be converted into
$\otimes$-twisted-$\epsilon$-good-pbits by LOCC, by prepending such
transformation to our scheme.
As an example, $S_e$ may contain tensor power of key distillable
states $\sigma^{\ot n}$ for arbitrarily large $n$.  Since $\sigma$ is
key distillable, $\exists k$ such that $\sigma^{\ot k}$ can be
preprocessed by ${\cal L}_k$ (via LOCC) to a state $\tilde{\sigma}_k$
that approximates some private state to some predetermined accuracy
(the dimension of the key part is then $2^{kr}$ for some $r>0$).
To test if $\untrust \in S_e$ (i.e., whether $\untrust = \sigma^{\ot
n}$), Alice and Bob can first apply the preprocessing ${\cal
L}_k^{\otimes \lfloor n/k \rfloor}$ to $\untrust$, followed by our
estimation procedure for $\tilde{S}_e$ containing
$\tilde{\sigma}_k^{\otimes \lfloor n/k \rfloor}$, which {\em is} a
$\otimes$-twisted-$\epsilon$-good-pbit.
Clearly if $\untrust \in S_e$, the above test will pass with high
probability.
There are several subtle points concerning this reduction:
(1) The key-rate can be suboptimal. (2) The preprocessing may prevent
the QKD scheme from being easily converted to \pmqkd schemes.  (3) The
dimension of the new key part, $2^{kr}$, is finite but can be large
for finite preprocessing precision, and the accuracy of our test has a
strong dimensional dependence.

% These problems may apply to other states that are LOCC-transformable
% to $\otimes$-twisted-$\epsilon$-good-pbits.
% 
% We will come back to discuss these problems.  
% 
We will also return to one other observation in Sec.\ \ref{sec:opu},
that for a given state, it can be related to many different pbits
(defined by different twisting operations).  Consequently, the error
rate of a state relative to each pbit and thus the key rate depend on
the choice of the pbit being considered, and should be optimized.  For
now, we consider an arbitrary choice, such as one arising from the
knowledge of the available channel.  Later, we will describe a simple
method for the optimization in Sec.\ \ref{sec:opu}.

Both \cite{HLLO05} and this paper exploit the relation between
$\epsilon$-good-ebits and $\epsilon$-good-pbits -- they differ only by
a change of basis (in particular, for
$\otimes$-twisted-$\epsilon$-good-pbits, the change is simply given by
the tensor power of the single-system twisting operation \eq{twist}).
That the twisting is not an LOCC operation, of course, changes all the
nonlocal resource accounting.
But surprisingly, as we will see, a variation of the Lo-Chau scheme is
invariant under twisting, except for one step.  So, we detail how and
why the Lo-Chau scheme works, and explain how that exceptional step
can be circumvened.

\subsection{Concepts in tolerable attacks} 
\label{sec:tolerableattacks}

Core to the analysis of QKD using noisy resources is a notion of
tolerable adversarial attacks, which are quantified by the parameters
to be estimated.  (For example, these are chosen to be the number of
bit-flip and phase-flip errors in the transmitted qubits in many
schemes.)
We make this notion precise in the following, and develop notations
used throughout the paper.
Consider an $n$-qubit system.
Let $\E$ be the Pauli group acting on it (parameter $n$ omitted).  For
each $P \in \E$, up to a scalar factor in $\{\pm 1,\pm i\}$, 
$P = \sigma_x^{x_1} \sigma_z^{z_1} \ot 
     \sigma_x^{x_2} \sigma_z^{z_2} \ot  \cdots 
     \sigma_x^{x_n} \sigma_z^{z_n}$ 
where $\sigma_{x,z}$ are the generators for the qubit Pauli group, and
$x_i, z_i \in \{0,1\}$ are matrix exponents.
It will become clear that the scalar factor is irrelevant in our work,
thus each $P$ is represented by the two $n$-bit strings ${\bf x} =
(x_1,x_2,\cdots,x_n)$ and ${\bf z} = (z_1,z_2,\cdots,z_n)$, which we
will call the ``$X$- and $Z$-components'' of $P$.
The number of $1$'s in a bitstring is called its Hamming weight.
Let $\e = (\e_x,\e_z)$.  They will represent two error rates critical
in the security of QKD.  
Collect all $P$'s in $\E$ that have $X$ and $Z$-components with
Hamming weights no greater than $n \e_x$ and $n \e_z$ into a set
$\E_\e$, and denote the linear span of $\E_{\e}$ (over $\CC$) by
$\SE_\e$.
The eavesdropping attack of current interest, described as a
trace-preserving completely-positive (TCP) map, is of the form
\be 
        {\cal P}_\e(\rho) = \sum_{k} E_k \, \rho E_k^\dagger
\label{eq:okerror}
\ee 
where $E_k \in \SE_\e$ for all $k$ and where the usual
trace-preserving condition $\sum_k E_k^\dagger E_k = I$ holds.
Note that $\e_x,\e_z \leq 1$, and when equality holds, $\SE_\e$ is the
set of all bounded operators, thus, {\em any} eavesdropping attack is
of the form \eq{okerror} for sufficiently large $\e_x,\e_z$.

For the case of qubit transmission, we omit the $d=2$ in the notation
for the maximally entangled state $|\Phi_d\>$ and $\Phi_d$.  Using 
\eq{okerror}, we make the important definition: 
\begin{definition}[$\e$-good-ebit] 
\label{def:ege} We call the state 
${\cal P}_\e(\Phi^{\otimes n})$ ``$n$ $\e$-good-ebits'', where ${\cal
P}_\e$ acts on the $n$ qubits of Bob and the identity map acts on the
$n$ qubits of Alice.
\end{definition}
Note that $\e$-good-ebits are {\em not} necessarily tensor power states.
We now define the analogue in a twisted basis: 
\begin{definition}[$\otimes$-twisted-$\e$-good-pbit] 
\label{def:egp} We call the state 
$U^{\ot n} [{\cal P}_\e(\Phi^{\otimes n}) \ot \rho_{\rm anc}]
U^{\dagger \ot n}$ ``$n$ $\otimes$-twisted-$\e$-good-pbits'', where $U$
is a twisting operator given by \eq{twist}, and the ancillary state
$\rho_{\rm anc}$ can be arbitrary over {\em all} the ancillary systems
$(A'B')^{\ot n}$.
\end{definition}

\subsection{The Lo-Chau scheme}  
\label{sec:lochau}

The Lo-Chau scheme focuses on the $d=2$ case.  Alice uses the channel
$n$ times to send Bob's halves of $n$ ebits she prepared locally.
In the absence of Eve, $\untrust$ shared after step (1) should differ
from $\Phi^{\ot n}$ by the channel noise.
It thus makes sense to use $\e = (\e_x, \e_z)$ as the parameter $e$ in
$S_e$, and define $S_\e$ to be the set of all $n$ $\e$-good-ebits. 
Here, $\e_x, \e_z$ are called the bit and phase error rates
respectively.
% 
% Note that we have intentionally omitted the system label for ${\cal
%  P}_{\e}$ since one can act on either system $A^{\ot n}$ or $B^{\ot
%  n}$ to produce the same state (similarly in \eq{lochau-pure-1}
%  below).
% 
When eavesdropping is possible, Alice and Bob need to determine $\e$ 
for which $\untrust \in S_\e$ with high probability.
In security proofs, we do not lose security if we assume less.  So, we
let $\untrust$ be completely unconstrained and allow Eve to possess
the purification of $\untrust$.  (A purification of a mixed state
$\rho$ on a system {\sc s1} is a pure state on two systems {\sc s1,s2}
such that tracing out the extra system {\sc s2} will give $\rho$.  The
purifying system {\sc s2} contains all information related to {\sc s1}
outside of it.)
Since $\{ P|\Phi\>_{AB}^{\ot n} \}_{P \in \E}$ is a basis for ${\cal
H}_{(AB)^{\ot n}}$, the purification of $\untrust$, $|\Psi_1\>$, has
the form
\be
        |\Psi_1\> = \sum_{P \in \E} \alpha_P  (P|\Phi\>_{AB}^{\ot n}) 
                    \otimes |e_P\>_E \,.
\label{eq:lochau-pure-1}
\ee
% 
% NB not a Schmidt decomposition
% 
Here, $P$ ranges over all possible $n$-qubit Pauli operators in $\E$
and acts on Bob's $n$ qubits, $\alpha_P$ are arbitrary amplitudes, and
$|e_P\>_E$ are normalized states on system $E$.

Step (2) in the Lo-Chau scheme is carried out by estimating the error
rates $\e_x$, $\e_z$ by {\em random sampling of the $AB$-systems without 
replacement}.
To estimate $\e_x$, $m$ systems are chosen randomly and $\sigma_z
\otimes \sigma_z$ is measured on each of them, and the estimated
$\e_x$ is the number of $-1$ outcomes divided by $m$.  (Note that the
outcome of measuring $\sigma_z \otimes \sigma_z$ on an ebit should be
$+(-)1$ when there is zero(one) $\sigma_x$ error).
In other words,
one measures $\lambda$, the eigenvalue of $\sum_{i=1}^{m} (\sigma_z
\otimes \sigma_z)_{AB}^{(i)}$ where $(i)$ denotes the $i$th sampled
system, and estimate $\e_x$ to be
$\smfrac{1}{2}(1-\smfrac{\lambda}{m})$.  (Similarly for $\e_z$.)
We will next describe the estimation process in two ways, a simple
abstraction and the actual implementation, and we show that they are
equivalent.

{\em In the abstract}, the error estimation transforms the state to
\be 
        \sum_P \alpha_P  (P|\Phi\>_{AB}^{\ot n}) \sum_{\e} 
                \mu_{P,\e} |\e\>_O  
                \otimes |e_P\>_E 
\label{eq:abstract} 
\ee
where the experimental estimate of the error rates of the QKD
execution, $|\e\> = |\e_x, \e_z\>$, is in the system $O$ available to
all three parties.
In a good estimation procedure, the estimated error rates should not
deviate significantly from the actual values, except with very small
probability.
Let $n \e_{xP}, n \e_{zP}$ be the Hamming weights of
the $X$- and $Z$-components of $P$.
A good estimation translates to the mathematical statement that, for
each $\e_{xP},\e_{zP}$, the sum of $|\mu_{P,\e}|^2$ over $\e$ should
be small whenever $|\e_{xP}-\e_x|$ or $|\e_{zP}-\e_z|$ is significant.
% 
% {\small no paragraph break 
% \\ BEGIN INTERNAL USE: Using p34 of \cite{LCA01} (and realizing that
% $A(\lambda,p) \approx (p{-}\lambda)^2/(p\,(1{-}p)\ln 2)$ to lowest order
% in $p{-}\lambda$ (so, $\lambda \leq p$ not necessary) and running the
% argument for both black and white balls, we get a $2$-sided bound.
% Let $m$ be sample size, $n$ be total number of systems, and use
% $\e_{xP},\e_{zP}$ and $\e_{x},\e_{z}$ as defined below.  
% % 
% \be
%       {\rm Pr}( |\e_{xP}{-}\e_{x}| \geq \delta) \leq 2 \cdot 
%       \exp \lpm -m (4 \delta^2 - \smfrac{m}{n-m})   \rpm  
% \ee
% % 
% If $m < (\smfrac{2\delta^2}{1{+}2\delta^2}) n$ (usually $m \sim
% \sqrt{n}$ or $\log n$), then,
% % 
% \be
%       {\rm Pr}(\, |\e_{xP}{-}\e_{x}| \geq \delta) \leq 2 \cdot 
%       \exp \lpm -2 m \delta^2 \rpm  
% \ee
% 
% END INTERNAL USE \\ continuing in the same paragraph} 
% 
% Using \cite{LCA01}, random sample (mod overlapping samples) can
% achieve the following.
% 
Reference \cite{LCA01} provides a test procedure for the Lo-Chau
scheme based on random sampling that achieves the following: For small
$\delta$ and for $m < (\smfrac{2\delta^2}{1{+}2\delta^2}) n$, we have
\bea
    {\rm Pr}( \, |\e_{xP}{-}\e_{x}| \geq \delta) \leq f(m,\delta)  
\label{eq:lca01}
\eea
where $f(m,\delta) := 2 \exp \lpm {-}2 m \delta^2 \rpm$ with
natural exp rather than base 2, 
% 
% dl: this should be changed in the next version 
% 
so that $|\mu_{P,\e_x,\e_z}|^2 \leq 2 f(m,\delta)$ if
$\max(|\e_x-\e_{xP}|,|\e_z - \e_{zP} |) \geq \delta$.  
In \eq{lca01}, the probability is over the random sample taken for 
the error estimate.
(See \cite{derive-lca01} for a derivation of \eq{lca01} from
\cite{LCA01}.)
% 
% this reference is between kraus-gisin-renner and ulhmann 
% 
To achieve good estimation is a central aspect of QKD.
The proof in \cite{LCA01} is subtle -- measurement of $P$ commutes
with measurement of $\e$ so that whether the former is done cannot
change the distribution of the latter.  So, we can assume measurement
of $P$ has been done here.
Most importantly, such assumption applies even to actual indirect
measurements of $\e$ that may not commute with measurement of $P$, as
long as the indirect measurement gives accurate results, and all
intermediate results (except for the final outcome) are discarded (see
argument to follow).
This imagined measurement of $P$ turns both $P$ and $\e$ into
classical random variables so that classical random sampling theory
can be applied.
% 
% This imagined measurement of $P$ is used in other parts of the
% security proof in a way similar to the above.  

{\em In real experiments}, there are two differences from the
abstraction.  First, the intended measurement operators $\sigma_x \ot
\sigma_x$ and $\sigma_z \ot \sigma_z$ are nonlocal (these are parity
measurements in the conjugate and computational bases) but they are
implemented via local measurements, for example, the eigenvalue of
$\sigma_x \ot \sigma_x$ is obtained by measuring that of $\sigma_x \ot
I$ and $I \ot \sigma_x$ {\em on the properly paired $AB$ systems} and
classically taking the product of the two outcomes ($\pm 1$).  Second,
the $2m$ random samples will be irreversibly measured out.
We want to replace the analysis of the real experiments by that of the
abstraction, and we now show that such replacement is valid if we
impose certain conditions on the protocol, as detailed in the
following observation:

\begin{obs}
If Alice and Bob perform (1) local (demolition) measurements, (2)
classical communication of the outcomes, (3) classical postprocessing
and output a function of the outcomes, (4) discard all measured
systems, all intermediate outcomes and communicated messages, then the
procedure is equivalent to a direct measurement yielding the
function (and nothing else) and discarding the measured system.

Proof: Local measurements can be made ``coherently'' so that the
outcome is stored in the computational basis of an ancilla without being
read.  Classical communication from Alice to Bob can be modeled as the
isometry $|x\>_A \rightarrow |x\>_A |x\>_B |x\>_E$ where Eve's copy
ensures classicality and generality of the security argument.
Similarly for classical communication from Bob to Alice.  Then, Alice
and Bob each performs the classical postprocessing (locally) to derive
the same intended measurement outcome.  Besides this, they discard
everything else, i.e.\, they give to Eve all measured systems, their
copies of the coherent classical communication, and the workspace of
the classical-post-processing, and Eve can reconstruct the
postmeasurement state.  The entire procedure is thus equivalent to the
desired direct measurement. 
\end{obs}

% {\em {\small Proof 2: Perhaps more obvious -- consider the
% estimation (including all symmetrization) as a measurement on all $n$
% systems.  If the actual one gives output distribution (output only
% $\e_x$, $\e_z$) similar to the ideal, and postmeasurement states are
% discarded, the two measurements are close to each other and quantum
% operations.  Proof by process-tomography.}} 

% {\em {\small This is put in so that the quantum-classical reduction will 
% fly pass in the private state case.} } 

Keep in mind not to use the $2m$ samples again, we can analyze the
state in \eq{abstract} in the abstract setting.  This state can be
rewritten as
\be 
        \sum_P \sum'_{\e} 
             \alpha_P \, \mu_{P,\e} \, (P|\Phi\>_{AB}^{\ot n})  
             \, |\e\>_O  
             \otimes |e_P\>_E 
       + |\Psi_{\rm bad}\>
\ee
where the primed sum of $\e$ is now restricted to those terms in which
$\e_x,\e_z$ are $\delta$ close to $\e_{xP},\e_{zP}$ respectively, and
the unnormalized $|\Psi_{\rm bad}\>$ contains all other terms with bad
estimates.  The important point is that $|\Psi_{\rm bad}\>$ has norm
squared bounded by $2f(m,\delta)$. (To see this, label the sum over
those $\e$ by a double prime, and $\sum_P \sum''_{\e} |\alpha_P
\mu_{P,\e}|^2 = \sum_P |\alpha_P|^2 \sum''_{\e} |\mu_{P,\e}|^2 \leq
\sum_P |\alpha_P|^2 2 f(m,\delta) \leq 2 f(m,\delta)$).  We include
this bad term in our equations to keep track of the entire picture but
we need not worry about its evolution.

In step (3), based on the estimates $\e_x, \e_z$, Alice and Bob run
any applicable EPP (e.g.\ see \cite{BDSW96,GL01}) on the unmeasured
$n-2m$ systems.
In the abstract, the state becomes 
\bea
\nonumber 
        \sum_P \sum'_{\e} \alpha_P \, \mu_{P,\e} \,  
             \lpm \beta_{\rm g} \, |\Phi\>_{AB}^{\ot (n-2m) r_\e} |{\rm g}_\e\>
                 + \beta_{\rm b} \, |{\rm b}_\e\>  \rpm  
             |\e \>_O  
             |e_P\>_E \\
        + |\Psi_{\rm bad}\> 
\nonumber
\eea
To obtain the above expression, note that the output of EPP depends on
$P$, $\e$, and {\em random inputs} of EPP.
Taking a coherent description for the local coins, and focusing on one
$(P,\e)$-term in the primed sum where error estimate is accurate, EPP
produces an output with high fidelity with respect to
$|\Phi\>_{AB}^{\ot (n-2m) r_\e}$ where $r_\e$ is the entanglement rate
depending mostly on $\e$ (and slightly on $n{-}2m$ for finite effect,
and finally, negligibly on the local coins because this effect can be
removed by lowering the rate slightly).
We collect the rest of the system into a sufficiently large auxiliary
space.  Uhlmann's theorem \cite{Uhlmann76} guarantees an output state
for the ($P$,$\e$)-term in the form inside the parenthesis: the
auxiliary output states $|\Phi\>_{AB}^{\ot (n-2m) r_\e} |{\rm g}_\e\>$
and the bad EPP term $|{\rm b}_\e\>$ are orthonormal, with $\beta_{\rm
g,b}>0$, $\beta_{\rm g}^2 + \beta_{\rm b}^2 = 1$ and $\beta_{\rm b}$
upper bounded by a function exponentially decaying with $n$
\cite{BDSW96,SP00,GL01,Hamada03}.
If $\e_x,\e_z$ are too high, $r_\e = 0$ and implicitly QKD is aborted 
(yet preserving security).
The incoherence between the different $P,\e$ terms can be absorbed
into the auxiliary system.
In real experiments, EPP is done incoherently, but as long as Alice
and Bob refrain from using anything other than the final output (i.e.\
discarding everything else) the abstract picture will hold.  

Finally, Alice and Bob measure out a key from the $(AB)^{\ot
(n-2m)r_\e}$ systems, which has high fidelity to ebits (when
conditioning on other systems is NOT made).  This guarantees security
\cite{BHLMO04} in the universal composable definition
\cite{BM02,BHLMO04}.
In particular, let the ideal state be $|\psi_{\rm ideal}\> =
\sum_{P,\e} \alpha_P \, \mu_{P,\e} \, |\Phi\>_{AB}^{\ot n} |{\rm
g}_{\e}\> |\e\>_O \ot |e_P\>_E$, and the output in the last equation
be $|\psi_{\rm actual}\>$.  Then, the QKD (in)security parameter in
\cite{BHLMO04} is upper bounded by $\sqrt{1- |\<\psi_{\rm
ideal}|\psi_{\rm actual}\>|^2} \leq \sqrt{4 f(\delta,m)+\beta_{\rm
b}^2}$ (because $|\<\psi_{\rm ideal}|\psi_{\rm actual}\>| \geq
(1-2f(\delta,m)) \beta_{\rm g}$).
Roughly speaking, it means that if an ideal key used in {\em any}
application is replaced by the one generated in the QKD protocol, no
attack involving all parts of the application can achieve a
statistical difference better than the stated insecurity parameter.

Note that if better EPP protocols are found and used in QKD (more
rapidly vanishing $\beta_{\rm b}$) the above analysis implies
corresponding improvement in the key rate and security of the
resulting QKD.  As a concrete example, \cite{GL01} presents a scheme
that achieves a key rate of $1-H(\e_x)-H(\e_z)$ where $H$ is the
binary entropy function and $\e_x,\e_z$ are the observed error rates,
and is nonzero for $\e_x < 1/2$ if $\e_z \approx 0$ (and vice versa)
or $\e_x = \e_z < 0.11$.  It follows from subsection \ref{sec:adv} 
that underlying states or channels with less error has the potential 
to establish a secure key. 

We end this section with a definition for a useful concept we came
across: 
\begin{definition}[Product vs nonproduct observables] 
\label{def:prodobs}
A product observable (with respect to systems S1, S2) is one of the
form $O_{S1} \otimes O_{S2}$.  While nonlocal, it can be measured
using LOCC: perform the individual local measurements $O_{S1} \otimes
I_{S2}$, $I_{S1} \otimes O_{S2}$, exchange the classical outcomes, 
and calculate the product. 
\end{definition}

\subsection{Replacing EPP by EC/PA}
\label{eppecpa}

In \cite{SP00,GL01}, classes of entanglement purification protocols
(EPP) were found to have a very nice property when used in EPP-QKD.
If Alice and Bob apply EPP followed by final measurements in the
computational basis to extract a key, their many steps can be
rearranged without changing the security of the final key.  In
particular, the rearranged protocol has the computational basis
measurement done first, generating what is called a ``sifted-raw-key''
(the adjective ``sifted'' is only useful later in the mapping to
P/M-QKD).  The steps of the original EPP become (classical) error
correction (EC) on the sifted-raw-key followed by privacy
amplification (PA) to generate the final key.  Such EPP include 1-EPP
protocols corresponding to CSS codes (i.e.\ involving only parity
checks entirely in the computational basis, or entirely in the
conjugate basis), and also 2-EPP protocols that are CSS like,
symmetric with respect to exchanging Alice and Bob, and with each step
depending only on prior measurement outcomes in the computational
basis.  We will call such schemes EC/PA-Lo-Chau schemes, which, from
now on, are always being considered in place of the original Lo-Chau
scheme.

References \cite{SP00,GL01,LCA01} provide recipes to convert EPP-QKD
to the simpler P/M-QKD.  We will first describe a pbit-distillation
based QKD scheme (PPP-QKD) and provide a security proof in the next
section.  In Sec.\ \ref{sec:pm}, we outline a conversion to \pmqkd for
our PPP-QKD scheme.

\subsection{QKD based on $\otimes$-twisted-$\epsilon$-good-pbits}
\label{sec:pstates}

We first consider the $d=2$ case in direct correspondence with the
EC/PA-Lo-Chau scheme, again omitting the $d=2$ notations in
$\gamma_d^U$ and $|\Phi_d\>$.

After step (1), Alice and Bob are sharing untrusted state $\untrust$,
and in step (2) Alice and Bob want to test if $\exists S_\e$ such that
$\untrust \in S_\e$, where 
\be S_\e = \{ U^{\ot n} \lpm {\cal P}_\e(\Phi_{AB}^{\ot n}) \ot
        \rho_{\rm anc}  \, \rpm  U^{\dagger \ot n} \} \ee
is a set of $\otimes$-twisted-$\epsilon$-good-pbits (see Def.\
\ref{def:egp}) for some $U$ satisfying \eq{twist} and some arbitrary
ancillary state $\rho_{\rm anc}$ on ${(A'B')^{\ot n}}$.  In principle,
Alice and Bob only need to find $\e$, but not $U$ and $\rho_{\rm anc}$
as long as they exist.  We will see that the protocol is independent
of $\rho_{\rm anc}$.  For now, we assume they make a certain guess for
$U$ and we will come back to remove this requirement in Section
\ref{sec:opu}.

Consider the following unfeasible scheme: first untwist, i.e., apply
$U^{\ot n \dagger}$ to $\untrust$, then apply the EC/PA-Lo-Chau scheme.
This is equivalent to running the EC/PA-Lo-Chau scheme in the case of
$S_\e = \{ {\cal P}_\e(\Phi_{AB}^{\ot n}) \}$ and thus it is secure.
The problem is that untwisting is global and requires resources
unavailable in real-life QKD.
Our strategy is to write down (mathematically) this secure but
unfeasible scheme as a first step.  Then, we explain
security-preserving modifications that make the scheme feasible using
the usual resources allowed in QKD.
In short, this is possible because only one step in EC/PA-Lo-Chau
scheme is affected by the twisting and untwisting operations (see also
\cite{HLLO05}).  The exceptional step is the estimation of $\e_z$ in
the twisted basis.  In \cite{HLLO05}, it was handled by first
distilling some ebits followed by teleportation of a small number of
test system to enable untwisting.  Here, it will be handled without 
distilling ebits.

In detail, this secure but unfeasible protocol runs as follows:

(2) Apply untwisting $U^{\ot n \dagger}$ to $\untrust$, then estimate
$\e_x$ and $\e_z$ on the $(AB)^{\ot n}$ systems (by using $m_x$ and
$m_z$ {\em random samples} respectively), and finally reapply $U^{\ot
  n}$.

(3') Apply untwisting $U^{\ot n \dagger}$, measure out a sifted-raw-key
in the $n{-}m_x{-}m_z$ systems.

(4') Perform error correction and privacy amplification on the
sifted-raw-key via $1$- or $2$-way public discussion.

We now explain how to transform the above protocol to one involving
only measurements of product observables and classical communication,
and in particular, without the distillation of ebits.
In step (2), only a random subset of $m_x{+}m_z$ systems are measured.  
On the other $n{-}m_x{-}m_z$ untested systems, the untwisting and
twisting cancel out (thus can be omitted).
On the tested systems, for the estimate of $\e_z$, untwisting,
measuring $\smfrac{1}{m_z} \sum_{i=1}^{m_z} (\sigma_x \ot
\sigma_x)^{(i)}_{AB}$ and twisting (where $i$ is a label of the tested
sample) is equivalent to the measurement of $\Gamma^{\rm ideal}_{\rm
\!x\,av} := \smfrac{1}{m_z} \sum_{i=1}^{m_z} \Gamma_x^{(i)}$ where
$\Gamma_x = U_{ABA'B'} (\sigma_x \ot \sigma_x \ot I_{A'B'})
U_{ABA'B'}^\dagger$.
Here, $\Gamma_x$ is a nonproduct observable (see
Def.~\ref{def:prodobs}) and generally, it cannot be measured in a
one-shot manner using LOCC.  However, our goal is to estimate $\e_z$
by measuring the {\em combined} $\Gamma^{\rm ideal}_{\rm \!x\,av}$, which
is an average of $\Gamma_x$ over many different systems, and for this
purpose, we can apply some other LOCC measurement, the method and the
accuracy will be given in the next paragraph.
For the estimate of $\e_x$, the twisted observable $\Gamma_z =
U_{ABA'B'} (\sigma_z \ot \sigma_z \ot I) U_{ABA'B'}^\dagger$ is simply
$\sigma_z \ot \sigma_z \ot I_{A'B'}$ because $\sigma_z$ commutes with
$U$ and $U^\dagger$.  So, the original analysis of \cite{LC99} holds,
and $m_x$ samples are used for this estimate.
In step (3') and (4'), the computational basis measurement to obtain the
sifted-raw-key and the rest of the classical postprocessing all
commute with $U^{\ot n \dagger}$.  Thus, we can finish the entire
QKD protocol before the untwisting, which then clearly does nothing
and can be omitted.

{\bf LOCC estimation of $\e_z$ via product observables:}

The goal is to replace a measurement of $\Gamma^{\rm ideal}_{\rm \!x\,av}
:= \frac{1}{m_z} \sum_{i=1}^{m_z} \Gamma_x^{(i)}$ by an LOCC
measurement of product observables, such that the outcomes have 
similar average values. 
%  
% such that the outcome statistics are similar.
% 
We denote the probability distribution of the outcome of measuring
$\Gamma^{\rm ideal}_{\rm \!x\,av}$ by $\mu_{\rm ideal}$, and that of the
LOCC measurement of product observables by $\mu_{\rm locc}$, and their 
averages by $\bar{\mu}_{\rm ideal}$ and $\bar{\mu}_{\rm locc}$.  
If the state being measured is fixed, $\mu_{\rm ideal}$ and
$\bar{\mu}_{\rm ideal}$ are fixed, but $\mu_{\rm locc}$ and
$\bar{\mu}_{\rm locc}$ are random variables depending on the
measurement outcomes.

We now explain the LOCC measurement that generates $\mu_{\rm locc}$.
First, obtain a decomposition for the single system observable
$\Gamma_x$ into product observables:
\bea 
	\Gamma_x 
	& = & U_{ABA'B'} (\sigma_x \ot \sigma_x \ot I) U_{ABA'B'}^\dagger
\\
	& = & \sum_{j_a,j_b=1}^{t} s_{j_a j_b} O_{j_a AA'} \otimes O_{j_b BB'}
\label{eq:gx}
\eea
where $\{O_{j}\}_{j=1}^t$ is a basis (trace-orthonormal) for hermitian
operators acting on $AA'$ and $BB'$, and $t = d^2 d'$.
Second, Alice and Bob divide their $m_z$ samples into $t^2$ groups.
They use each group for one pair of $(j_a,j_b)$, and they obtain a
measurement outcome denoted by 
${\rm Out}_{m_z/t^2}[O_{j_a AA'} \otimes O_{j_b BB'}]$ 
of the observable $\frac{t^2}{m_z} \sum_{i=1}^{m_z/t^2} (O_{j_a AA'}
\otimes O_{j_b BB'})^{(i)}$.  This is related to a sum of product
observables, and can be measured in LOCC as mentioned before (Alice
and Bob can individually measure $O_{j_a AA'}$ and $O_{j_b BB'}$ on
the $i$th test system, multiply their results via LOCC, and finally
sum those products over $i=1,\cdots,m_z/t^2$) and take average.
Also, let $\sum_{j_a,j_b} s_{j_a j_b} {\rm Out}_{m_z/t^2}[O_{j_a AA'}
\otimes O_{j_b BB'}]$ be the ``outcome'' of the LOCC estimation of the
phase error rate, defining a distribution $\mu_{\rm locc}$.  

Is $\bar{\mu}_{\rm locc}$ close to $\bar{\mu}_{\rm ideal}$ that is
generated by measuring $\Gamma^{\rm ideal}_{\rm \!x\,av}$ directly?
It will be if the entire $m_z$ sample systems are in a joint
tensor-power state, and if $m_z/t^2$ is large enough (because
Chernoff-like bounds will hold and because of \eq{gx}).
However, in our current problem, Alice and Bob share $\untrust$ which
is {\em not} a tensor-power state.  Fortunately, first, by means of
random sampling, we can assume permutation symmetry in this analysis,
and second, since the estimation involves only a small portion ($m_z$)
of the entire $n$ systems, the exponential quantum de Finetti theorem
\cite{Renner05} states that the measured (reduced) state is close to a
mixture of ``almost-tensor-power-states'' so that the Chernoff-like
bounds will hold and the estimate will thus be accurate.
The exact analysis involves many adaptations of the results in
\cite{Renner05}.
In the appendix, we prove a more general theorem (Theorem
\ref{th:box2}) for any observable $O$ on one copy of the bipartite
Hilbert space in place of $\Gamma_x$, for any dimensions, and for
$\mu_{\rm ideal}$ generated by measuring ${1 \over m} \sum_{i=1}^m
O^{(i)}$ and $\mu_{\rm locc}$ generated by the above LOCC procedure.
We obtain an upper bound for $\Pr (|\bar{\mu}_{\rm ideal} -
\bar{\mu}_{\rm locc} | > \delta)$ as follows.   

Adapting Theorem \ref{th:box2}, we write the symbol in the appendix on
the left hand side of the arrow, and what it should be in the current
context on the right hand side.  We choose the parameters as
$d\rightarrow d^2 d'$, $t \rightarrow t^2 = d^4 d'^2$,
$n+2m\rightarrow n$, $m \rightarrow m_z$, $\delta \rightarrow
\delta/3$, $\| \Gamma_x \|_{HS}^2 = d^2 d'$ (since $\Gamma_x$ is
unitary).  Then,
\bea 
	& & \Pr (\,|\bar{\mu}_{\rm ideal} - \bar{\mu}_{\rm locc}| > \delta) 
\nonumber \\ 
  &\leq & 2 e^{-\smfrac{(n-m_z)(r+1)}{2n} + \smfrac{1}{2} d^4 d'^2 \ln (n-m_z)}
\nonumber \\ 
  &+& (t^2+1) 2^{-\left[\smfrac{\delta^2}{36 t^2 d^2 d'} 
                   - H(\smfrac{rt^2}{m_z})\right] \smfrac{m_z}{t^2} 
		   + d' d^2 \log(\smfrac{m_z}{2t^2}+1)}
\nonumber \\ 
  &+& 2 e^{-\smfrac{m_z \delta^2}{144 d' d^2}}
\label{eq:distdiff}
\eea
where the three expressions in the upper bound respectively come from
the exponential quantum de Finetti theorem, the Chernoff bound, and
random sampling theory.  (In the last term of the above, we have used
a tighter bound given directly by Proposition \ref{prop:srodka}
instead of the general bound in Theorem \ref{th:box2}.)  Also,
throughout the paper, $H(\cdot)$ denotes the binary entropy function.

Furthermore, by the sampling theory Proposition \ref{prop:srodka} in
the Appendix, $\e_x$ can also be estimated with $m_x$ samples to the
accuracy ${\rm Pr}(| \, \e_{x,P} - \e_x | > \delta) \leq 2
e^{-\smfrac{m_x \delta^2}{16}}$.  

Putting these altogether, ${\rm Pr}(| \, \e_{z,P} - \e_z | \geq \delta
{\rm ~or~} | \, \e_{x,P} - \e_x | \geq \delta) \leq f(m_x,m_z,\delta)$
for
\bea 
& &  f(m_x,m_z,\delta) \nonumber \\
& \leq & 2 e^{-m_x \delta^2/16}
\nonumber \\ 
  &+& 2 e^{-\smfrac{(n-m_z)(r+1)}{2n} + \smfrac{1}{2} d^4 d'^2 \ln (n-m_z)}
\nonumber \\ 
  &+& (t^2+1) 2^{-\left[\smfrac{\delta^2}{36t^2 d^2 d'} 
                   - H(\smfrac{rt^2}{m_z})\right] \smfrac{m_z}{t^2} 
		   + d' d^2 \log(\smfrac{m_z}{2t^2}+1)}
\nonumber \\ 
  &+& 2 e^{-\smfrac{m_z \delta^2}{144 d' d^2}}
\label{eq:finalf}
\eea
The composable security parameter will still be less than $\sqrt{4
f(\delta,m_x,m_z)+\beta_{\rm b}^2}$ as derived in the summary of the
original Lo-Chau scheme.  
 
Now, we state parameters that will make {\small $\sqrt{4
f(\delta,m_x,m_z)+\beta_{\rm b}^2}$} exponentially small in some
security parameter $s$.  Note that $\beta_{\rm b}$ is unaffected by
our modification to the EC/PA-Lo-Chau scheme, and we focus on the
$f(\delta,m_x,m_z)$ portion.  We choose some security parameter $s$
and make each term in \eq{finalf} exponentially small in $s$ ($\approx
2^{-s}$) by the following choices (with each item corresponding to
each term in order). \\
(1) Take sample size for bit-error rate to be $m_x = s \times
    \smfrac{16}{\delta^2}$. \\
(2) Generally, since $m_z$ has to be small compared to $n$, thus, $r =
    4 s$ and $r \geq d^4 d'^2 \ln n$. \\
(3) $m' := m_z/t^2$ should be large (at least $O(s)$), while $r/m' \ll
    1$ and $m' \gg \log m'$. In particular, say, $H(r/m') \leq
    \delta^{2}/(72 t^2 d^2 d')$ and $m' \delta^2/ (72 t^2 d^2 d') \geq
    2 d' d^2 \log(m'/2+1)$ and $m' \geq s \times 144 t^2 d^2 d' /
    \delta^2 - 2 \log t$.\\
(4) $m' \geq \smfrac{s{+}1}{t^2} \smfrac{144 d' d^2}{\delta^2}$.  \\
Clearly, for $s$ ranging from constant to linear in $n$, there are
corresponding choices of $r,m_z,m_x$ that will work.  Roughly
speaking, $m_x \geq O(s/\delta^2)$ and $m_z \geq O(\log n)$ will be
the two asymptotic requirements when $d,d'$ are fixed.
The final key rate can be given by $(1-(m_x+m_z)/n) \times
R_{\e_x,\e_z}$ where the first factor is due to the use of private
states and the resulting more complicated error estimation procedure
(but lower bounded by the above choices of $m_x$ and $m_z$) and the
second factor depends on the observed {\em twisted} error rates (that
can be much lower than that relative to ebits and where our protocol
provides an advantage) and the choice of EPP (or EC/PA procedure) 
lower bounds of which are extensively studied in QKD based on ebits.

With this analysis of the accuracy of the estimation, and following
from earlier discussion, the security proof for the QKD protocol is
completed.  

% wcl: too vague
% to make it clear, it is very restrictive and not very useful
% to leave it unclear, there are other parts of QKD that can kill the 
% statement.  

%%%%%%%%%%%%%%%%%%%%%%%%%%%%%%%%%%%%%%%%%%%%%%%%%%%%%%%%%%%%%%%%%%%%%%%
% We finish this discussion with the following observation: 
% 
% \begin{obs}
% 
% Our method to twist the EC/PA-Lo-Chau scheme can be applied to other
% analoguous QKD schemes are also secure.  The key rates as functions of
% the estimated parameters, are the same in the twisted and untwisted
% scenario.
% \end{obs}
%%%%%%%%%%%%%%%%%%%%%%%%%%%%%%%%%%%%%%%%%%%%%%%%%%%%%%%%%%%%%%%%%%%%%%%

\section{Prepare and measure scheme} 
\label{sec:pm}

In the previous section, we have provided a security proof of the
pbit-purification-based QKD (PPP-QKD) protocols in which the parties
are processing an untrusted shared state and are extracting a key from
it.  Typically, the processing requires quantum memory, and some
times, coherent operations on the quantum state.
As we have noted, entanglement-distillation-based protocol (EPP-QKD)
are closely related to the much simpler P/M-QKD.
% 
% DL: we say these TOOO many times 
% 
% in which Alice prepares and sends a quantum state to Bob (through an
% untrusted channel) who measures it immediately.
% % 
% \pmqkd requires very little quantum coherent manipulation and are much
% simpler to implement.  With this in mind, 
We will thus convert our PPP-QKD protocol to a \pmqkd scheme, adapting
to pbits earlier works based on ebits \cite{SP00,LCA01,GL01}.  

In PPP-QKD, the initial state is completely arbitrary.  In P/M-QKD,
Alice first prepares the state, and then Eve attacks it.  Thus, the
state $\untrust$ is more restricted.  In particular, we focus on
tensor power states prepared by Alice -- the most physically relevant
case because of the simplicity in implementation.

Since our protocol already has the distillation steps replaced by 
EC/PA, there are only $2$ coherent steps to modify: 
(1) distribute $\untrust$ via an untrusted channel and 
(2) estimate $\e_x,\e_z$ on the sample systems and measure the rest in
the computational basis to generate the sifted-raw-key. 

We now dissect these two steps.  
In most of the useful cases, in step (1), Alice only needs to prepare
a tensor power state $\initial^{\otimes n}$ over the $n$ bipartite
systems and send each of Bob's halves via one use of the given
untrusted channel ${\cal N}$.
They are expecting to share the state $({\cal I}\otimes {\cal
N}(\initial))^{\otimes n}$ while they are actually sharing $\untrust =
({\cal I}^{\otimes n} \otimes {\cal E}(\initial^{\otimes n}))$ for an
arbitrary joint attack ${\cal E}$ by Eve.
For step (2), recall that it suffices to perform measurements of
product observables on individual system.
Now, focus on such a measurement of some $O_a \otimes O_b$ on one of
these systems.  Let $\{|\psi_l\>\}_l$ be a complete set of
eigenvectors of $O_a$.
Note that Alice's measurement on each of her halves of the state {\em
commutes} with the transmission via the channel and Bob's measurement.
So, she can measure first, before sending each of Bob's halves,
without affecting the security.
It means that she sends the state ${\rm
tr}_{AA'}[(|\psi_l\>\<\psi_l|_{AA'} \otimes I_{BB'}) \, \initial
\,\!_{AA'BB'}]$ (unnormalized) with probability which is the trace of
that state, followed by a measurement of $O_b$ by Bob.
Thus, a conversion to \pmqkd can be obtained, with a caveat.  

The problem is that in purification-based QKD, each pair of local
measurements for each system is chosen probabilistically (from Eve's
point of view) and with perfect coordination between Alice and Bob.
When converting to P/M-QKD, various naive options fail or come with 
extra requirements: 
(a) If Alice announces her measurement before Bob signals receipt of
the states, Eve could have intercepted the transmitted state,
performed Bob's measurement, resent the postmeasurement state to Bob,
and completely evaded detection.  
(b) If Alice waits until Bob signals receipt of the states, before
announcing her basis, and then Bob makes his measurements, he will
need quantum memory to hold his received states.
(c) If Alice's bases annnouncement is encrypted with a private key, it
has to be of length roughly $O(m_z \log n) = O(s\,{\rm
polylog}(n)/\delta^2)$ where $s$ is the security parameter of the QKD
protocol (which can range from constant to linear in $n$).
In comparison, an initial key is also required for the authentication
of some of the classical messages.  It is an open problem what is the
minimum authentication requirement.  If one authenticates all of the
bases information, the identity of the states in the test samples for
parameter estimation, and forward communication in EC/PA, it will take
$O(\log n)$ key bits.
% 
% The total size of this message is
% $\approx O(m_z \log n) + O(d \sqrt{d'} (m_x+m_z)) + O(n)....$
% 
Thus, for high security parameter requiring $m_z$ to be growing with
$n$, encryption of the bases information qualitatively increases the
the amount of the initial key required.

The initial solution in BB84 was to have Bob guessing the measurement
basis, and postselect those with properly matched measurement basis.
The price is a lower key rate.
The method was improved on by \cite{LCA01} so as to preserve the
key rate asymptotically.  We will adapt this technique in our protocol.

The idea in \cite{LCA01} is that, even though randomness in the
measurement basis is necessary for security, only a small fraction
needs to differ from the computational basis to have sufficient
confidence in the estimates of $\e_x,\e_z$ -- something we have
already exploited in the PPP-QKD scheme in the previous section.
Here, Alice and Bob will independently pick a large enough fraction
$O(n^{c})$ of the $n$ systems to be measured for each $O_j$ (the
orthonormal basis for operators acting on each of the local systems
$AA'$ and $BB'$).  These samples are chosen {\em randomly} and with
high probability over the choice, for each pair $(j_a,j_b)$, the
observable $O_{j_a} \otimes O_{j_b}$ would have been applied to a
fraction $O(n^{2c-2})$ of all systems, giving $O(n^{2c-1})$ random
samples.  Remember the requirement $m_z \geq s \, {\rm
polylog}(n)/\delta^2$, so that $n^{c} \geq \sqrt{s n} \, {\rm
polylog}(n)/\delta$ will provide sufficient overlap for calculcating
$\< O_{j_a AA'} \otimes O_{j_b BB'} \>$ and subsequently $\e_z$.
For the protocol in the previous section, the local dimensions are $d
\sqrt{d'}$ and each of Alice and Bob have $t = d^2 d' - 1$ traceless
observables to measure locally.  Thus $O(t n^{c}) \approx o(n)$
systems will be used for estimating $\e_z$ and the rest can all be
measured in the computational basis (for estimating $\e_x$ and for the
(unsifted) raw-key generation) thus the key rate of the original
PPP-QKD scheme is preserved.  Finally, by the procedure to turn a
measurement of Alice and $\initial$ into an ensemble of signal states,
the conversion to a \pmqkd is completed.

We note that the above procedure can be suboptimal, especially if $t$
is large.  For example, if the decomposition of the single system
observable $\Gamma_x$ has low Schmidt rank $t' \ll t^2$ in the
Hilbert-Schmidt decomposition, then, effectively, only $t'$ local
observables have to be measured.  Also, data from unmatching bases can
be potentially useful but in the current scheme they are discarded for
simplicity of the analysis.  These, and other optimization, are issues
for future research.

\section{A channel with zero quantum capacity and nonzero key rate}
\label{sec:example}

Recall that there are key distillable but bound entangled states
\cite{HHHO03,huge-key,HPHH05}.  Using results in the current paper,
they can be verified and therefore can have nonzero rates of
generating unconditionally secure key.  The adversarial setting is
totally unconditional, as described in Subsection \ref{sec:adv}.
Based on one of these states, we construct a channel that has zero
quantum capacity and nonzero key rate.

The channel is defined as follows.  According to Section \ref{sec:pm},
Alice prepares a tensor power state $\initial^{\otimes n}$ over $n$
bipartite systems and sends each of Bob's halves via one use of the
given untrusted channel ${\cal N}$.
They are expecting to share the state $({\cal I}\otimes {\cal
N}(\initial))^{\otimes n}$ while they are actually sharing $\untrust =
({\cal I}^{\otimes n} \otimes {\cal E}(\initial^{\otimes n}))$ for an
arbitrary joint attack ${\cal E}$ by Eve.
We choose ${\cal I}\otimes {\cal N}(\initial)$ to be $\rho_H$ from
\cite{HPHH05} (the definition will be given later).  This state has
$3$ desirable properties. (1) $\rho_H$ has a maximally mixed reduced
state on $AA'$.  Thus it can indeed be written as $ \rho_H = ({\cal
I}_{AA'} \otimes {\cal N}_{BB'})(\Phi_{dAB} \otimes \Phi_{d'A'B'})$
for some channel ${\cal N}_{BB'}$.
% 
%(As an aside, it also has the advantage of being low-dimensional, with
%input/output dimension of ${\cal N}$ being $4$.)  
% 
(2) $\rho_H$ is PPT (having positive partial transpose \cite{Peres96})
and is thus bound entangled.  
Since $\rho_H$ is bound entangled if and only if ${\cal N}$ is
entanglement binding (with zero rate to create entanglement for any
unentangled input) (\cite{HHH99}), ${\cal N}$ has zero quantum
capacity.
(3) On the other hand, if verified, $\rho_H$ has nonzero key rate.
Correspondingly, in the absence of eavesdropping, Alice and Bob can
use ${\cal N}$ to distribute copies of $\rho_H$, verify them, and
generate a key.  
Thus $\rho_H$ and ${\cal N}$ provide the example we are seeking.

% Amusingly, the error rates associated with ${\cal N}$ is as high as
% $\e_x=1/2$ and $\e_z=1/(1+\sqrt{2})$ on the first input qubits and
% $\e_x=\e_z=1/2$ on the second, giving an average total error rate of
% 95.7\%.  (FIX: DOUBLE CHECK)

We now define the state $\rho_H$.  Recall that for a pure state
$|\psi\>$, we use the shorthand $\psi$ for the density matrix
$|\psi\>\<\psi|$.
Define the four Bell states as
\ben
|\psi_{0,1}\> = {1\over \sqrt{2}}(|00\> \pm |11\>) \\
|\psi_{2,3}\> = {1\over \sqrt{2}}(|01\> \pm |01\>)
\een
with the projectors given by $\psi_i$. 
Define also the states
\be
	|\chi_\pm\> ={1\over 2} (\sqrt{2 \pm  \sqrt{2}}\,\, |00\> 
	\pm \sqrt{2 \mp \sqrt{2}}\,\,|11\> )
\ee
Then, for $\kappa$ a small parameter to be defined later, take 
\be
\rho_H = (1-\kappa) \sum_{i} q_i \psi_{iAB} \ot \rho_{A'B'}^{(i)}
	+ \kappa \, \smfrac{I}{16} 
\label{eq:rhoh}
\ee
where $q_0 = q_1 = \frac{p}{2}$, $q_2 = q_3 = \frac{1{-}p}{2}$, and 
\ben
&&\rho^{(0)} = {1\over 2}[|00\>\<00| + \psi_2] \nonumber  \\
&&\rho^{(1)} = {1\over 2}[|11\>\<11| + \psi_3] \nonumber \\
&&\rho^{(2,3)} = \chi_{+,-}
\een
For $p = {\sqrt{2}\over 1 + \sqrt{2}}$, and $\kappa = 0$, $\rho_H =
\rho_H^{T_2}$ where $T_2$ denotes the partial transpose of the second
system.   In particular, $\rho_H$ is PPT.  
When $p \neq {\sqrt{2}\over 1 + \sqrt{2}}$, the $1{-}\kappa$ term need
not be PPT anymore, but choosing a small nonzero $\kappa$ will give a
corresponding neighborhood of $p$ for which $\rho_H$ will remain PPT
and thus bound entangled.
Here, we claim that there is an untwisting operation of the form given
by \eq{twist} that we can apply to $\rho_H$, so that further tracing
of $A'B'$ will give us the state
\be
	\sigma_{AB} = (1-\kappa) \lpm p \psi_0 + (1-p) \psi_2 \rpm  
	+ \kappa \smfrac{I}{4}
\,.  
\label{eq:untwistednoisypbit}
\ee 
Note that the transformation of the $\kappa$ term is straightforward
and also $\kappa \rightarrow 0$, so, we can focus on the $(1-\kappa)$
term.
Let $V_{1}$ be any unitary that transforms the following states as: 
\bea
	|00\> \ra |00\> \,,  
	|11\> \ra |11\> \,,  
	|\psi_2\> \ra |01\> \,,  
	|\psi_3\> \ra |10\> \, 
\eea
and $V_2$ transforms $|\chi_{+,-}\>$ to $|00\>,|11\>$ respectively.  
$V_{1,2}$ exist, because they preserve orthonormality of the input 
space. 
Then, the untwisting operation can be defined as: 
\bea
	U_H & = & \lpm |11\>\<11|_{AB} \otimes \sigma_{zA'} \ot I_{B'}\rpm 
	\times 
\nonumber \\
	      & & \lpm (|00\>\<00|+|11\>\<11|)_{AB} \ot V_{1 A'B'}
\nonumber \\
	      & &    + (|01\>\<01|+|10\>\<10|)_{AB} \ot V_{2 A'B'} \rpm 
\eea
This is because the right hand factor first transforms the
$\rho^{(0,1,2,3)}_{A'B'}$ component in $\rho_H$ to
$\smfrac{1}{2} (|00\>\<00|+|01\>\<01|)$, 
$\smfrac{1}{2} (|11\>\<11|+|10\>\<10|)$, 
$|00\>\<00|$, and $|11\>\<11|$, respectively, 
and the subsequent left hand factor (effectively a
controlled-$\sigma_z$ from $A'$ to $A$) turns $\psi_{1,3}$ into
$\psi_{0,2}$ respectively.  This proves the assertion that the
untwisted state is given by \eq{untwistednoisypbit}.
The exact key rate in an execution will depend on the observed error
rates $\e_x,\e_z$, but it can potentially be close to being given by
\eq{untwistednoisypbit}.
In this case, for very small $\kappa$ and $p > 1/2$ (we have taken $p
\approx 0.5858$), certain EPP (1-way asymmetric CSS EPP) studied in
\cite{BDSW96,GL01} can distill a key from $\untrust$ at a rate
$1-H(\e_x)-H(\e_z)$ which is $\approx 1-H(0.5858) \approx 0.0213$
according to \eq{untwistednoisypbit}.  Thus there exists a
corresponding EC/PA procedure to generate a key from our protocol.
This completes the proof that an untrusted channel, supposedly ${\cal
N}$, is entanglement binding but can have nonzero key rate.

As a side remark, the first term of $\rho_H$ represents as mixture of
two pbits with a common twisting operation (no constraint on ancillary
state) but differing by a bit-flip of the underlying ebit.  EPP on
these mixtures are particularly simple.

Let us also understand this channel ${\cal N}$ in more operational
terms.  Recall that a channel is completely determined by the state
${\cal I} \ot {\cal N}(\Phi_d)$ where $d$ is the input dimension of
${\cal N}$.  Thus, any correct way to transform $\Phi_{AB} \ot
\Phi_{A'B'}$ to $\rho_H$ via operations on $BB'$ will be a valid
description of how ${\cal N}$ acts.
Thus consider the following sequence of operations: \\
(1) With probability $p/2$, measured $B'$ in the computational basis.
(a) If outcome is $|0\>$, do nothing.  
(b) If outcome is $|1\>$, apply $\sigma_z$ to $B$. \\
(2) With probability $p/4$, apply $\sigma_{zB} \otimes \sigma_{yB'}$. \\
(3) With probability $p/4$, apply $\sigma_{xB'}$. \\
(4) With probability $(1-p)$, measure $B'$ with the following POVM
\begin{eqnarray}
M_0  =  \left[ \begin{array}{cc} 
\frac{1}{2}(\sqrt{2+\sqrt{2}}) & 0 \\ 
0 & \frac{1}{2}(\sqrt{2-\sqrt{2}}) \end{array} \right]; \\
M_1  =  \left[ \begin{array}{cc} 
 \frac{1}{2}(\sqrt{2-\sqrt{2}})  & 0 \\ 0 
&\frac{1}{2}(\sqrt{2+\sqrt{2}}) \end{array} \right].
\end{eqnarray}
(a) If the outcome is ``0'', apply $\sigma_{xB}$. 
(b) If the outcome is ``1'', apply $\sigma_{yB} \otimes \sigma_{zB'}$. 

It is immediate that cases (1a), (1b), (2), (3), (4a), and (4b) give
post-measurement states (systems labeled by $ABA'B'$):
$\psi_0 \ot |00\>\<00|$, 
$\psi_1 \ot |11\>\<11|$, 
$\psi_1 \ot \psi_3$, 
$\psi_0 \ot \psi_2$, 
$\psi_2 \ot \chi_+$, 
and $\psi_3 \ot \chi_-$, 
respectively. 
Also, both measurements yield equiprobable outcomes.  
Thus, the probabilities for all these cases are $p/4,p/4,p/4,
p/4,(1-p)/2,(1-p)/2$.
Mixing up the states from all the cases gives exactly the $(1-\kappa)$
term of $\rho_H$.  To incorporate the negligible $\kappa$ term, we can
take convex combination of the above with the completely randomizing
channel on $BB'$.
While the bit and phase error rates are tricky to define, any simple
attempt will yield amusingly high numbers.

Finally, we can use the recipe in Sec.\ \ref{sec:pm} to obtain a
corresponding P/M-QKD scheme.  
The initial state $\initial$ in this case is $\Phi_{AB} \ot
\Phi_{A'B'}$. 
For the estimation of $\e_z$, Alice will measure product Pauli
operators on $AA'$ because they form the desired orthonormal basis for
traceless observables.
The actual states transmitted are exactly the equiprobable ensemble of
the six eigenstates of $\sigma_{x,y,z}$ on each of $B$ and $B'$ for
the $O(\sqrt{ns})$ test systems.
The rest of the systems are prepared in random states in the
computational basis.
Likewise, Bob measures $O(\sqrt{ns})$ systems in the eigenbases of
$\sigma_{x,y,z}$ and the rest in computational basis.
Note that the ensemble of states sent and the measurements on Bob's
side are exactly those of the efficient version of the six-state
protocol.  Here, a completely different interpretation via pbits, and
a corrrespondingly different classical postprocessing scheme yield 
drastically different results (the key rate will be zero otherwise).

\section{Summary of protocols} 
\label{sec:sumprot}

We summary our PPP-QKD protocol and the P/M-QKD protocol in the
following.  We do not repeat why it is secure and omit the security 
parameters that are covered in Secs.\ \ref{sec:details} and \ref{sec:pm}. 

We refer to $n$ copies of bipartitite systems $AA'~BB'$
with $\dim(AA'BB') = D$.  Alice and Bob preagree on some product basis
of $D^2$ Hermitian operators $O_{j_a} \otimes O_{j_b}$.  Let $O_1$ be
diagonal in the computational basis.  Throughout, random sampling of
the systems is done without replacement.

\vspace*{1ex} 

A general PPP-QKD protocol will proceed as follows:

(1) Alice and Bob share untrusted states $\untrust$ using their
    underlying quantum resources.  The state is supported on $n$
    bipartite systems $AA'~BB'$, with .
    
(2) (a) Alice and Bob jointly pick $m_x$ systems at random (by using
    local coins and $1$-way public discussion) and independently
    measure the $A$ and $B$ parts in the computational basis, and
    combine the outcomes to obtain the observed error rate $\e_x$.
    (b) For each $j_a,j_b$, they jointly pick $m_z/D^2)$ sample
    systems and measure $O_{j_a}$ on $AA'$ and $O_{j_b}$ on $BB'$
    independently.  Then, for each candidate untwisting operation,
    they combine the outcomes to obtain $\e_z$.  They pick the lowest
    value.  They measure the rest of the $A$ and $B$ systems in the
    computational basis.

(3) Based on $\e_z$ and $\e_x$, they apply an appriopriate EC/PA 
    procedure.  

(4) They obtain a key of rate determined by $\e_z$ and $\e_x$. 

Step (2b) is the only placing differing from the standard
EC/PA-Lo-Chau schemes. 

\vspace*{1ex}

A general P/M-QKD has the following form: 

(1) Alice and Bob share some underlying untrusted channel ${\cal N}$
acting on $BB'$ and can use it $n$ times.  They agree on some $\rho_0$
supported on system $AA'BB'$.  

(a) For each $j_a$, let $|\psi_l\>_{AA'}$ be the eigenvectors of
$O_{j_a}$, and Alice transmits a state ${\rm tr}_{AA'} [
(|\psi_l\>\<\psi_l|_{AA'} \otimes I_{BB'}) \rho_{AA'BB'}$
(renormalized) via one use of the channel (without knowing what
happens to the actual transmission) .  This state is labeled by $j_a$
and $l$, and let the normalization be $p^{j_a}(l)$.  The $(j_a,l)$
state is transmitted via $n^c p^{j_a}(l)$ randomly chosen uses of the
channel.  The rest of the channel uses is the same but always has 
$j_a = 1$.

(b) For each $j_b$, Bob measures $O_{j_b}$ on $n^c$ randomly chosen
channel outputs.  The rest are measured in $O_1$.  

(2) Alice and Bob then start public discussion.  They use systems
transmitted based on $O_{j_a}$ and measured based on $O_{j_b}$ to
calculate the average of $O_{j_a} \otimes O_{j_b}$ (the average of
$ll'$ where Alice transmits state labeled by $l$ and Bob's outcome is
$l'$).

(a) They obtain a direct estimate of $\e_x$.

(b) For each candidate untwisting operation, they obtain an estimate
of $\e_z$, and they pick the lowest value. 

(3) EC/PA is applied. 

(4) A key is generated. 

\section{Optimal untwisting} 
\label{sec:opu}

In our QKD protocol (both the purification-based and the P/M variant),
the key rate is determined by the estimated $\e_x,\e_z$, and once
these are measured, we optimize over the EPP or the EC/PA procedure.

Consider, for each $U$, the PPP-QKD scheme again.  Given $\untrust$,
we want to find $\e_x,\e_z$ such that $\untrust \in S_\e = \{ U^{\ot
n} ({\cal P}_\e(\Phi_{AB}) \ot \rho_{A'B'}^{\ot n}) U^{\dagger \ot n}
\}$ and generally, such $\e_x,\e_z$ will depend on $U$.
As long as $\untrust \in S_\e$ for some $U$, $\e$ is a legitimate
estimate and EPP will produce a secure key of appropriate length.
(Untwisting only occurs in our interpretation of the sampled data.)
To exploit this feature, note that more precisely, $\e_x$ is
independent of twisting, but $\e_z$ is not.
Thus, for a list of possible twisting operators $U_i$, Alice and Bob
should estimate each of the corresponding twisted phase error rate
$\e_{zi}$ and take the minimal one to optimize the key rate
extractable in EPP.
At a first glance, they will need to measure $\Gamma^{\rm ideal}_{\rm
\!x\,av}$ for each $U_i$.  But recall that each twisted phase error is
derived from the decomposition given by \eq{gx} and from estimating
the product observables in the decomposition.  For different $U_i$,
the same set of product observables are measured, and the detail on
$U_i$ only enters the QKD protocol in the coefficients in the
decomposition \eq{gx}, and thus, the same set of product observables
can be used to calculate all possible $\e_{zi}$, and the optimization
over twisting operator is an entirely classical computation problem.

Similar analysis holds for P/M-QKD.  Just like PPP-QKD, the choice of
$U_i$ only enters the protocol via the classical computation of the
estimate $\e_{zi}$.  Thus, Alice and Bob runs the protocol as stated
before, but now with extra minimization of $\e_{zi}$ over all possible
$U_i$ in their classical computation, followed by the appropriate
EC/PA procedure.

\section{Discussion} 
\label{sec:discussion}

We have seen that for any channel which allows for the distribution of
key distillable states, there exists a protocol for verifying
security.  The protocol is related to the scheme of Lo and Chau, the
difference being that phase errors become twisted phase errors, and
are measured by decomposing this operator in terms of product
observables.  Accuracy of this procedure is due to the exponential
quantum de Finetti theorem \cite{Renner05} and the usual Chernoff bound
and sampling theory.
Security of this protocol was proven by reduction to the Lo-Chau
proof of security.  We then converted it to a prepare and measure
scheme which has the advantage of not requiring quantum coherent
control.  Furthermore, one can classically optimize over the twisting
operation to minimize the corresponding twisted phase error rate, and
thus maximize the key rate.
More generally, each EPP-QKD protocol 
that involves parameter estimation on a small fraction of sample
systems and only computational basis measurement and classical
processing of the data
has a PPP-QKD analogue and a P/M-QKD analogue.
Paradoxically, though the heart of the security proof relates to
entanglement purification, it never needs to be done in the actual
protocol, nor is noiseless entanglement needed in our scheme. 
In particular, our protocol can be based on bound entangled states or
binding entanglement channels with zero entanglement rate or quantum
capacity.

This demonstrates conceptually that quantum key distribution is not
equivalent to the ability to send quantum information.  
The ``information gain implies disturbance'' effect is strong enough
to provide security even in such noisy regime.  It also means that the
ability to perform near perfect error correction on any logical space
is unnecessary.

As a side result, the procedure outlined in the Appendix can be of
independent interesting -- in particular, it follows that the average
value of an observable $O$ on a large number of systems (without any
underlying structure) can be estimated by measuring a sublinear sample
even if $O$ has to be measured indirectly in terms of a decomposition
to other observables (in our case product observables) that do not
necessary commute with $O$.

We have noted that some of the states or measurement results are not
used in the analysis.  Further research will exploit such data to
improve on the key rate.

It will also be interesting to study the alternative protocol based on
state tomography discussed in Sec.\ \ref{sec:related} and
\cite{HHHO03} and investigate possible advantages (such as that on the
key rate).

A big open question is whether all entangled states can be converted
via LOCC to pbits, and related to this question is whether all binding
entanglement channels can be used for QKD.

Finally, our protocol is restricted to some classes of twisting (e.g.,
tensor power twisting).  It will be interesting to either show the
possibility of QKD in the case of completely arbitrary twisting or to
obtain a no-go theorem.

\appendix

\section{LOCC estimation of the expectation of an IID observable}
%convex combination of such states along different vectors $\theta$.

\subsection{Finite quantum de Finetti theorem and generalized Chernoff bound}

We say that a state $\rho_n$ on Hilbert space $\hcal^{\ot n}$
satisfies the Chernoff bound with respect to a state $\sigma$ on
$\hcal$ and a measurement $\mcal$ on $\hcal$ if (with high
probability) the {\em relative frequency distribution} obtained by
measuring $\mcal^{\ot n}$ on $\rho_n$ is close to that of measuring
$\mcal$ on $\sigma$.  For example, $\rho_n = \sigma^{\ot n}$.
However many other states satisfy the same property.  An important
class is called {\em almost power states}, which are formulated and
studied in \cite{Renner05}.  We adapt results in \cite{Renner05} for our
own purpose in the following.

\begin{theorem}{\bf (Finite quantum de Finetti theorem plus Chernoff bound)}
\label{thm:fincher}
Consider any permutationally invariant (possibly mixed) state
$\rho_{n+k}$ on Hilbert space $\hcal^{\ot (n+k)}$. Let $\rho_n=\tr_k
\rho_{n+k}$ be the partial trace of $\rho_{n+k}$ over $k$ systems.
Let $0 \leq r \leq n/2$.  Then there exists a probability measure
$\mu$ on (possibly mixed) states $\sigma$ acting on $\hcal$ and a
family of states $\rho^{(\sigma)}_{n,r}$ such that
\bee
\item The state $\rho_n$ is close to a mixture of the states
$\rho^{(\sigma)}_{n,r}$
\be
\label{ineqfin}
\left\| \rho_n - \int \rho^{(\sigma)}_{n,r} \; 
{d}\mu(\sigma) \right\|_{\rm tr} \leq 2\, 
e^{-{k(r+1)\over 2(n+k)} + {1\over 2} \dim(\hcal)^2 \ln k }
\ee
\item The states $\rho^{(\sigma)}_{n,r}$ (called {\rm almost power states}) 
satisfy the Chernoff bound in the following sense
\ben
\label{ineqcher}
{\rm Pr} \left( \left \|
P_{\mcal}(\sigma) - Q_{\mcal}[\rho^{(\sigma)}_{n,r}]\right \| > \delta 
\right) 
\nonumber \\ \leq 
2^{-n \, \left[ {\delta^2 \over 4} - H({r\over n}) \right] 
+ |W|\log ({\smfrac{n}{2} + 1})} =:
e(\delta)
%
% DL: use \te tilde e not to confuse with e in the main text.
%
\een
where $\mcal{=}\{M_w\}_{w\in W}$ is any measurement on $\hcal$,
$P_{\mcal}(\sigma) = \{{\rm Tr}(\sigma M_w)\}_w$, 
$Q_\mcal[\rho^{(\sigma)}_{n,r}]$ is the frequency distribution obtained
from measuring $\mcal^{\ot n}$ on the state
$\rho^{(\sigma)}_{n,r}$, and $|W|$ is the size of the alphabet $W$.
\item Reduced density matrices of the states $\rho^{(\sigma)}_{n,r}$
(to $n' \leq n$ systems with $r \leq n'/2$) satisfy the same Chernoff bound:
\ben
{\rm Pr}( \left \| 
P_{\mcal}(\sigma) - Q'_{\mcal}[\rho^{(\sigma)}_{n,r,n'}] \right \| > \delta) 
\nonumber \\ \leq 
2^{-n' \, \left[ {\delta^2 \over 4} - H({r\over n'}) \right] 
+ |W|\log ({\smfrac{n'}{2} + 1})}
\een
where $\rho^{(\sigma)}_{n,r,n'}=\tr_{n-n'} \rho^{(\sigma)}_{n,r}$ is
the resulting state after partial tracing $n-n'$ systems from
$\rho^{(\sigma)}_{n,r}$ and $Q'_\mcal[\rho^{(\sigma)}_{n,r,n'}]$ is
the frequency distribution obtained from measuring $\mcal^{\ot n'}$ on
the state $\rho^{(\sigma)}_{n,r,n'}$.
\eee 
Throughout the theorem, the probability is taken over the actual
measurement outcomes that defines the frequency distributions.  We
also use $[\cdot]$ for frequency distributions defined by measurement
outcomes whenever appriopriate.
\end{theorem}

\noindent {\em Proof:} We first collect various facts, definitions,
and results from \cite{Renner05}.

\subsubsection{Fact and definitions} 

{\definition {\bf Almost power state: (Def.\ 4.1.4, in \cite{Renner05})}
Suppose $0 \leq r\leq n$.  
Let $\; \Sym(\hcal^{\ot n})$ denote the symmetric
subspace of pure states of Hilbert space $\hcal^{\ot n}$.  
Let $|\theta \> \in \hcal$ be an arbitrary pure state and consider: 
\bea
\vcal(\hcal^{\otimes n}, |\theta\>^{\ot n - r}) := 
\{\pi(|\theta\>^{\ot n-r}\ot |\psi_r\>): 
\pi \in S_n, \, 
\nonumber 
\\|\psi_r\> \in \hcal^{\ot r} \}
\nonumber 
\eea
where $S_n$ is the permutation group of the $n$ systems.
We define the {\it almost power states along $|\theta\>$} to be
the set of pure states in 
\be 
|\theta\>^{[\otimes,n,r]} 
% :=  Sym(\hcal^{\ot n},|\theta\>^{\ot n - r})
:= \Sym(\hcal^{\ot n}) \cap 
{\rm span}(\vcal(\hcal^{\ot n}, |\theta\>^{\ot n - r}))
\ee
We denote the set of mixtures of {\it almost tensor power states}
along $|\theta\>$ as ${\rm conv}(|\theta\>^{[\otimes,n,r]})$.}

\noindent With the above definition, we shall prove the following lemma: 

{\lemma ~~ 
If $\varrho_{n} \in {\rm conv}(|\theta\>^{[\otimes,n,r]})$, then,
$\varrho_{n-m} \in {\rm conv}(|\theta\>^{[\otimes,n-m,r]})$ where
$\varrho_{n-m}={\rm Tr}_m(\varrho_{n})$ is the reduced density matrix
after the partial trace over any $m$ out of the $n$ systems (by
symmetry, without loss of generality, we take the first $m$ systems).
\label{PSym}
}

{\it Proof .-} 

Since membership in ${\rm conv}(|\theta\>^{[\otimes,n-m,r]})$ is
preserved under mixing, it suffices to prove the lemma for pure
$\varrho_{n} = |\Psi_n\>\<\Psi_n|$, with $|\Psi_n\> \in 
|\theta\>^{[\otimes,n,r]}$.

We can pick an ensemble realizing $\varrho_{n-m}$ of our choice, and
prove the lemma by showing that any element $|\Psi_{n-m}\>$ in that
ensemble belongs to $|\theta\>^{[\otimes,n-m,r]}$.
Our ensemble is obtained by an explicit partial trace of $|\Psi_{n}\>$
over the first $m$ subsystems along the computational basis.  An element 
is given by
\be
|\Psi_{n-m}\rangle=\<i_{1}|...\<i_{m}| \otimes I_{n-m}|\Psi_{n}\rangle.
\label{AnsambleVector}
\ee
Now, we note two facts: 

(i) $|\Psi_{n-m}\> \in \Sym(\hcal^{\otimes(n-m)})$, since
$|\Psi_{n}\rangle \in \Sym(\hcal^{\otimes n})$.
% DL: don't need the following
% = {\rm span}(|\phi\>^{\otimes n})$.

(ii) $|\Psi_{n-m}\> \in \vcal(\hcal^{\otimes n{-}m},
|\theta\rangle^{\otimes (n-m-r)})$ -- This is because $|\Psi_{n}\> \in
\vcal(\hcal^{\otimes n}, |\theta\>^{\ot (n{-}r)})$, and expressing
$|\Psi_{n}\>$ in terms of the spanning vectors of
$\vcal(\hcal^{\otimes n}, |\theta\>^{\ot (n{-}r)})$ and putting it into
Eq.~(\ref{AnsambleVector}), we have
\be
|\Psi_{n-m}\rangle = \sum_{\Psi_{r},\pi} \alpha_{\Psi_{\!r},\pi} 
 \<i_{1}| \cdots \<i_{m}| \otimes I_{n-m} \; {\pi}  \, 
(|\theta\rangle^{\otimes (n{-}r)} \otimes
 |\Psi_{r}\rangle).
\nonumber
\ee
Elementary analysis shows that any term of the above sum is, up to
permutation, of the form {\small 
 $(\langle i_{1}|\theta \rangle) \cdots (\langle i_{p}|\theta \rangle) 
|\theta\>^{\otimes n-r-p} \otimes
[\langle i_{p+1}| \cdots \langle i_{m}| \otimes I_{r-(m-p)} \; \pi'  
(|\Psi_{r}\rangle)] $ }
where $0 \leq p \leq m$, and ``absorbing'' $m-p$ copies of $\theta$ to
the last part of the vector, we get
$|\theta\>^{\otimes n-(r+m)} \otimes |\Psi''_{r}\>$.  
Thus, $|\Psi_{m-n}\>$ is a sum of terms of the form
$\pi(|\theta\>^{\ot n-(r+m)} \otimes |\Psi''_{r}\>)$, and belongs to
$\vcal(\hcal^{\otimes (n-m)}, |\theta\>^{\ot n - (r+m)})$.  
This proves the second fact, and also the lemma.  $\square$

The next lemma asserts that a mixture of almost tensor power states
behaves approximately like a mixture of tensor power states with
respect to a generalized version of Chernoff bound.

{\lemma {\bf (Theorem 4.5.2 of \cite{Renner05})} 
Let $0 \leq r \leq \frac{n}{2}$, $|\theta\> \in \hcal$, and
$|\Psi_{n}\>$ be a vector from $|\theta\>^{[\otimes,n,r]}$.
Let ${\cal M} = \{M_w\}_{w\in {\cal W}}$ be a POVM on ${\hcal}$, 
$P_{\mcal}(|\theta\rangle\langle\theta|)$ be the probability
distribution generated by applying the measurement to
$|\theta\rangle\langle\theta|$ (i.e.,
$P_{\mcal}(|\theta\rangle\langle\theta|) = \{{\rm Tr} |\theta \rangle
\langle \theta|M_w\}_w$), and $P_{\mcal}[|\Psi_{n}\rangle\langle
\Psi_{n}|]$ be the relative frequency distribution of outcomes of
${\cal M}^{\ot n}$ applied to $|\Psi_{n}\rangle \langle \Psi_{n}|$.
Then, 
\ben
\Pr \lpm \, \| P_{\mcal}(|\theta\rangle \langle \theta|) -
P_{\mcal}[|\Psi_{n}\rangle \langle \Psi_{n}|] \, \| > \delta \rpm \nonumber
\\\leq 2^{-n \, \left[ {\delta^2 \over 4} - H({r\over n}) \right] 
+ |W|\log ({\smfrac{n}{2} + 1})}
\nonumber =: e(\delta)
\een
where the probability is taken over the outcomes.  Note that we have
used $e(\delta)$ instead of $\delta(e)$ in \cite{Renner05}.
\label{GenChernoff}
}
 
Consider the general probability ${\rm Pr}( \| P_{\mcal}(\rho) -
P_{\mcal}[\varrho_{n}] \|<\delta)$ where $P_{\mcal}[\varrho_{n}]$ is a
frequency distribution of outcomes of ${\cal M}^{\ot n}$ applied to
$|\Psi_{n}\rangle \langle \Psi_{n}|$.  The distribution
$P_{\mcal}[\varrho_{n}]$, if treated as a functional of $\varrho_{n}$
on the space $\hcal^{\ot n}$, is {\it linear} in $\varrho_{n}$.
Following this we get immediately: 

\bec Lemma \ref{GenChernoff} holds when replacing the projector
$|\Psi_{n}\rangle \langle \Psi_{n}|$ (for $|\Psi_{n}\> \in
|\theta\>^{[\otimes,n,r]}$) by $\varrho_{n} \in {\rm
conv}(|\theta\>^{[\ot, n,r]})$.
\label{GenChernoffMixed}
\eec
 
Apart form the generalized Chernoff-type lemmas, we also need the
crucial exponential quantum finite de Finetti theorem:

\begin{theorem}[Theorem 4.3.2 of \cite{Renner05}]
For any pure state $|\psi_{n+k}\> \in \Sym(\hcal^{\ot n+k})$ and
$0\leq r\leq n$ there is a measure $d\nu (|\theta\>)$ on $\hcal$ and
for each $|\theta\> \in \hcal$ there is a pure state
$|\psi_n^{(\theta)}\> \in |\theta\>^{[\ot, n,r]}$ such that
\ben
& & \left\| 
{\rm Tr}_k |\psi_{n+k}\>\<\psi_{n+k}| -  \int_{\hcal} 
|\psi_n^{(\theta)}\>\<\psi_n^{(\theta)}|  d\nu (|\theta\>)
\right\|_{\rm tr} 
\nonumber \\
& \leq & 2e^{-{k(r+1)\over 2(n+k)} + {1\over 2} \dim(\hcal)\ln k } 
\label{ineqfor_convhull}
\een
\label{th:for_convhull}
\end{theorem}
Finally, we need the fact that any permutationally invariant 
state has a symmetric purification.

\bel[Lemma 4.2.2 of \cite{Renner05}]
\label{lem:symmpurif}
Let $\rho_n$ be a permutationally invariant state on $\hcal^{\otimes
n}$.  Then there exists a purification of the state in $\Sym((\hcal\ot
\hcal)^{\ot n})$.  \eel

This concludes the list of facts and definitions needed for proving 
Theorem \ref{thm:fincher}.  

\subsubsection{Proof of Theorem \ref{thm:fincher}}

Consider an arbitrary permutationally invariant state $\varrho_{n+k}$ on
Hilbert space $\hcal^{\ot (n+k)}$.  \\
Step (1): According to Lemma
\ref{lem:symmpurif} there is a purification $|\psi_{n+k}\>$ that
belongs to $\Sym(\hcal'^{\ot n+k})$ where $\hcal'=\hcal \otimes
\tilde{\hcal}$ and dim$(\tilde{\hcal})=$dim$(\hcal)$. \\
Step (2): We apply Theorem \ref{th:for_convhull} to $|\psi_{n+k}\>$ with
the changes 
\bea
\hcal & \rightarrow & \hcal'=\hcal \otimes \tilde{\hcal}
\nonumber
\\
d & \rightarrow & d^2
\eea
Step (3): After application of theorem \ref{th:for_convhull} we perform
partial trace over $\tilde{\hcal}^{\ot n}$, the purifying space 
introduced in step (1).  We denote this partial trace by $\tilde{\rm Tr}$.
This partial trace induces from the measure for pure state on $\hcal'$
in step (2) a new measure $\mu(\sigma)$ on the set of all mixed states
$\sigma$ acting on $\hcal$.  (The probability of $\sigma$ is given by
the total probability of all $|\theta\>$ with $\tilde{\rm
Tr}(|\theta\rangle\langle \theta|)=\sigma$).  This partial trace
produces also the states $\almpower$ defined directly by $\almpower
\equiv \tilde{\rm Tr}(|\psi_n^{(\theta)}\rangle \langle
\psi_n^{(\theta)}|)$ where the existence of the pure states
$|\psi_n^{(\theta)}\>$ is guaranteed by Theorem \ref{th:for_convhull}.
Finally we note that partial trace does not increase the trace
distance between two quantum states, so applying partial trace to the
LHS of (\ref{ineqfor_convhull}) and using the notation described above
we get immediately the inequality (\ref{ineqfin}).  This proves the
first item of Theorem (\ref{thm:fincher}).

To prove the second item of Theorem (\ref{thm:fincher}), remember from
the above that $\almpower \equiv \tilde{\rm
Tr}(|\psi_n^{(\theta)}\rangle \langle \psi_n^{(\theta)}|)$.  Since
$|\psi_n^{(\theta)}\rangle$ is an almost power pure state, lemma
\ref{GenChernoff} applies.  Further, it holds for all POVM-s on
$\hcal'=\hcal\ot\tilde{\hcal}$, and in particular for incomplete
POVM-s acting only on $\hcal$ but not on $\tilde{\hcal}$.  Thus, the
conclusion of lemma \ref{GenChernoff} holds with the change: $\mcal
\rightarrow \mcal \otimes I$, which gives item (2).  

Finally, to prove item 3 of theorem \ref{thm:fincher}, note that the
reduced density matrices $\varrho_{n,r,n'}^{\sigma}$ of interest can be
obtained from the pure state $|\psi_n^{(\theta)}\>$ above by tracing
(i) first over $n{-}n'$ subsystems corresponding to $\hcal'$, producing a
state on $\hcal'^{\ot n'}$, and
(ii) then over $n'$ subsystems corresponding to $\tilde{\hcal}$.

Then lemma \ref{PSym} guarantees that the first partial trace produces
a mixed state $\varrho_{n'}$ in ${\rm
conv}(|\theta\>^{[\ot,n',n'-r]})$ (with underlying space $\hcal'$.
Applying corollary \ref{GenChernoffMixed} to $\varrho_{n'}$ with $n'$
instead of $n$, it suffices to consider a pure state in
$|\theta\>^{[\ot,n',n'-r]}$.  Finally, lemma \ref{GenChernoff} can be
applied to this pure state with $\mcal \rightarrow \mcal \otimes I$
which concludes item 3. $\square$

\subsection{Two other useful results} 

\subsubsection{Classical random sampling} 

In addition to the fact and definitions above and Theorem
\ref{thm:fincher}, we will need the following result on classical
random sampling (or equivalently symmetric probability distribution). 

\bep{\bf(Classical sampling theory) Lemma A.4 from \cite{RK04b}.}
\label{prop:srodka}
Let $Z$ be an $n$-tuple and $Z'$ a $k$-tuple of random variables over
 a set $\zcal$, with symmetric joint probability $P_{ZZ'}$. Let
 $Q_{z'}$ be the relative frequency distribution of a fixed sequence
 $z'$ and $Q_{(z,z')}$ be the relative frequency distribution of a
 sequence $(z,z')$, drawn according to $P_{ZZ'}$. Then for every $\ep
 \geq 0$ we have \be \Pr_{Z\!Z'}(||Q_{(z,z')} - Q_{z'}||\geq \ep )\leq |\zcal|
 \, e^{-{k\ep^2/8|\zcal|}}.  \ee \eep The result says that the relative
 frequency distribution obtained from a small sample is close to one
 obtained from the whole system.  (This lemma is similar to
 \eq{lca01}, but stronger in two respects -- it applies to any
 dimension and has no restriction on the fraction sampled.  On the
 other hand, \eq{lca01} has better constants in the exponent.)

%%%%%%%%%%%%%%%%%%%%%%%%%%%%%%%%%%%%%%%%%%%%
\subsubsection{From probabilities to averages}
%%%%%%%%%%%%%%%%%%%%%%%%%%%%%%%%%%%%%%%%%%%

{\lemma Consider an observable $L$ on Hilbert space $\hcal$,
$\dim\hcal=d$. Let $L=\sum_{i=1}^{t} s_i L_i$, where $\{L_i\}$ is a
trace orthonormal basis for operators (i.e., $\tr L_i
L_j^\dagger=\delta_{ij}$).  Let the eigenvalues of $L_i$ be denoted by
$\lambda_l^{(i)}$.  Consider an arbitrary state $\rho$, and let
$P^{(i)}=\{p_l^{(i)}\}$ be the probability distribution on $l$ (which
eigenvalue) induced by measuring $L_i$ on $\rho$.  Let
$Q^{(i)}=\{q_l^{(i)}\}$ be an arbitrary family of distributions on
the eigenvalues of $L_i$.  We then have
\bea
\nonumber
& & \left| \<L\>_{\rho} - \sum_{i} s_i \sum_l \lambda_l^{(i)}q^{(i)}_l 
\right| \\
& \leq & 
\sqrt{t} \left\| L \right\|_{HS} \max_i \| P^{(i)}-Q^{(i)} \|_{\rm tr} \, 
\eea
where $\| \cdot \|_{HS}$ is the Hilbert-Schmidt norm and $\| \cdot
\|_{\rm tr}$ is the trace norm.  
\label{closeaverages}
}

{\it Proof}
\ben
&& \left|
\<L\>_{\rho} - \sum_{i} s_i \sum_l \lambda_l^{(i)} q^{(i)}_l \right|
\nonumber
\\
&=&
\left|
\sum_i s_i \sum_l \lambda_l^{(i)}(p_l^{(i)} {-} q_l^{(i)}) \right| 
\nonumber
\\
&\leq& 
\sum_i |s_i| \; \lpm \max_l |\lambda_l^{(i)}| \rpm \, 
\| P^{(i)}-Q^{(i)} \|_{\rm tr} 
\nonumber
\\
&= & 
\sum_i |s_i| \; \| L_i \|_{\infty} \, \| P^{(i)}-Q^{(i)} \|_{\rm tr}
\nonumber
\\
& \leq & 
\lpm \max_i \| P^{(i)}-Q^{(i)} \|_{\rm tr} \rpm 
 \sum_j |s_j| \; \| L_j \|_{\infty} 
\een
where $\| \cdot \|_\infty$ is the operator norm.
Since $||L_i||_{\infty} \leq 1$, using convexity of $x^2$ we obtain 
\be
\sum_{j=1}^t \, |s_j| \, \| L_j \|_{\infty}
\leq \sum_j |s_j| \leq \sqrt{t} \sqrt{\sum_j s_j^2}=\sqrt{t} \, \|L\|_{HS}
\ee
This completes the proof. $\square$ 
% 

%%%%%%%%%%%%%%%%%%%%%%%%%%%%%%%%%%%%%%%%%%%%%%
\subsection{Estimation - detailed description}
%%%%%%%%%%%%%%%%%%%%%%%%%%%%%%%%%%%%%%%%%%%%

We consider $2m+n$ systems with Hilbert space $\hcal^{\ot {(2m +n)}}$,
$\dim \hcal = d$ in a permutationally invariant state $\varrho_{2m+n}$.
%
% We then consider fixed measurement on these systems. For given state
% of these systems, the measurement induces a joint probability
% distributions on its outcomes.
%
Suppose the ultimate goal is to obtain the ``empirical mean-value'' of
some single-system observable $\Sigma$ on a sample of $n+m$ systems. 
In other words, we want to measure
$\frac{1}{N}\sum_{j=1}^{N}\Sigma^{(j)}$ where $\Sigma^{(j)}=I \otimes
I \otimes \cdots \otimes \Sigma \otimes \cdots \otimes I$ on the $N$
subsystems for $N=n+m$.
 
Because of experimental limitations (here, it is the LOCC constraints
on Alice and Bob), they are restricted to measuring product operators of the
form $L = L_A \otimes L_B$ by independently finding the eigenvalues of
$L_A$ and $L_B$ (i.e., making the measurements $L_A \otimes I$ and $I
\otimes L_B$), discussing over classical channels and multiplying
their outcomes together.  
Now, to measure $\Sigma$, one can first rewrite it in terms of 
product operators $L_i$: 
\begin{equation}
\Sigma=\sum_{i=1}^{t} s_{i} L_{i}
\label{eq:local_decomp}
\end{equation}
where we have chosen $\{L_i\}$ to be hermitian and trace orthonormal, so 
that $s_i$ are real.  The $L_i$-s are ``{\em intermediate observables}.''
We will describe an inference scheme that (1) involves only the
estimation of the ``empirical mean-value'' of $\Sigma$ on a small
number ($m$) of subsystems, and (2) the measurement of $\Sigma$ is
done indirectly via measurements of the $L_i$'s. 

The analysis will start with a special assumption about the $2m$-element
sample, $m$ of which are used for indirect estimation.  The
assumptions are relaxed on that sample.  After that properties of the
other $m+n$ subsystems are inferred.

%%%%%%%%%%%%%%%%%%%%%%%%%%%%%%%%%%%%%%%%%%%%%%%%%%%%%%%%%%%%%%%%%%%%
\subsubsection{Analysis of the $2m$ sample 
in an ``almost power state along $\sigma$'':$\Romr$}
%%%%%%%%%%%%%%%%%%%%%%%%%%%%%%%%%%%%%%%%%%%%%%%%%%%%%%%%%%%%%%%%%%%%
 
Suppose the first $2m$ subsystems are in a joint state $\Romr$, with
$r\leq \frac{1}{2} \times 2m$.  We expect the state $\Romr$ to play a
role similar to the state $\sigma^{\ot 2m}$.
Define the theoretical direct average 
\begin{equation}
\Sgm={\rm Tr}(\Sigma \sigma)=\sum_{i}s_{i}\langle L_{i}\rangle_{\sigma}
\label{eq:ss}
\end{equation}
We now consider an indirect measurement of $\Sigma$ applied on the
first $m$ subsystems, and a direct measurement on the next $m$
subsystems.
% 
% It should be clear enough that the direct and indirect measurements 
% are on different parts of the joint system.  
%
We will show that the empirical average, either obtained directly or
indirectly, will be close to the above.  

For the indirect measurement, divide the first $m$ subsystems into
$t$ groups. Each group has $m'=m/t$ subsystems.
Alice and Bob take the $i$th group ($i=1,\cdots,t$) and measure $L_i$ on each
site as described above (the measurement is ${\cal L}_{i}$).
In other words, the measurement $M^{\rm indirect} = \otimes_{i=1}^{t}(
{\cal L}_{i}^{\otimes m'})$ is applied to the first $m$ subsystems of
the entire $2m+n$ subsystems.
% 
%The reduction of the state $\Romr$ to the first $m$ subsystems induces
%a probability distribution $\pcal$ on the outcomes of $M^{\rm
%indirect}$.

We expect $\Romr$ and $\sigma^{\ot m}$ to behave similarly.  In
particular, consider an observable $L_{i}=\sum_{l}\lambda_{l}^{(i)}
|\psi_l^{(i)}\>\<\psi_l^{(i)}|$ expressed in its spectral
decomposition and the probability distribution on the set of
eigenvalues $\acal_i$ induced by the state $\sigma$ as follows:
\begin{equation}
P_{i}=\{ {\rm Tr}(\sigma |\psi_l^{(i)}\>\<\psi_l^{(i)}| ) \}_l
\end{equation} 
An execution of the measurement ${\cal L}_{i}^{\otimes m'}$ gives a
particular outcome $(l_{1},...,l_{m'})$ and induces a relative
frequency distribution $Q_{i}$ on $\acal_i$.

Then, the empirical frequency distributions $Q_i$ is close to the
``theoretical'' distribution $P_i$: 
\begin{fact}
\label{fact:distributions}
\be
 \Pr(\| P_{i}-Q_{i} \|_{\rm tr} \geq \delta)\leq e(\delta,m',r,d),
\ee
where the probability is taken over the measurement outcomes, $d$ is
the dimension of the single site Hilbert space, and
\begin{equation}
 e(\delta,n,r,|{\cal Z}|) \; 
 {:}{=} \; 2^{-(\frac{\delta^{2}}{4}-H(\frac{r}{n}))n +
 |{\cal Z}| \log(\frac{n}{2}+1)}
\end{equation}
\end{fact}

{\it Proof -} Follows immediately from the third item of Theorem
\ref{thm:fincher}.  Note that we use item (3) not (2) since we perform
the measurement only on {\it part} of the state $\Romr$.

{\it Remark -} Note also that $P_{i}$ is constant while $Q_{i}$
is a random variable.

Now, we define the theoretical average values for the intermediate 
observables $L_i$'s:
\be
\Li = {\rm Tr}(L_{i}\sigma)
\ee
and the empirical averages 
\be 
\Liemp=\sum_{l}\lambda_{l}^{(i)}Q_{i}(l)
\ee
where $Q_i(l)$ denotes the value of $Q_i$ on a specific event $l$ in
the alphabet $\acal_i$. (Again, $ \langle L_{i} \rangle_{\sigma}$ is
constant while $\Liemp$ is a random variable
depending on the particular outcomes of the measurements, and recall
that $L_{i}=\sum_{i}\lambda_{l}^{(i)}|\psi_l^{(i)}\>\<\psi_l^{(i)}|$).
Denote the {\it empirical} value of $\Sigma$ obtained {\it
indirectly} {\it via} the empirical averages of the $L_{i}$'s by 
\begin{equation}
\SgmBemp = \sum_{i}s_{i} \Liemp  \,.
\end{equation}
We now show that the indirect empirical average is close to the 
direct theoretical average in \eq{ss}.  
First applying the union bound to Fact \ref{fact:distributions}, we get
\begin{equation}
\Pr(\cup_{i=1,...,t}\{ \| P_{i} -Q_{i} \|_{\rm tr}>\delta  \})\leq
t\cdot e(\delta,m',r,d)
\end{equation} 
Then using Lemma \ref{closeaverages} we obtain that
\bea
\label{Ineq1}
& & \Pr \left( \left| \Sgm-\SgmBemp \right| > \delta \right) \nonumber \\
& = & \Pr \left( \left| 
\sum_{i}^{t}s_{i}\Li -\sum_{i}^{t}s_{i}\Liemp \right |>\delta
\right)
\nonumber \\
& \leq & t \cdot
e \left( \smfrac{\delta}{ \| \Sigma \|_{HS}\sqrt{t}},m',r,d \right) \,.
\eea
We emphasize once again that the probabilities are taken over 
the measurement outcomes. 

After considering the indirect measurements, suppose that someone
measures directly $M^{\rm direct} = \smfrac{1}{m} \sum_{j=m+1}^{2m}
\Sigma^{(j)}$ on the second group of $m$ subsystems.
Denote the empirical average outcome by $\SgmAemp$. 
In a way similar to the indirect case (but much easier here) we show 
that the empirical direct average is close to $\Sgm$ in \eq{ss} 
(by applying Lemma \ref{closeaverages} with $t=1$): 
\be
\label{Ineq2}
\Pr \left( \left| \Sgm - \SgmAemp \right| >\delta \right) \leq
e \left( \smfrac{\delta}{ \|\Sigma \|_{HS}},m,r,d \right) \,.  
\ee

{From} the inequalities (\ref{Ineq1}), (\ref{Ineq2}) we obtain
\bel
\label{lem:SigmaDistance}
For the measurements on the state $\Romr$ 
considered above we have:
\bea
& & \Pr \left(\left|\SgmAemp-\SgmBemp\right|>2 \delta\right) \nonumber 
\\ & \leq & t \cdot e\left(\smfrac{\delta}
{\| \Sigma \|_{HS}\sqrt{t}},m',r,d\right) +
e\left(\smfrac{\delta}{\|\Sigma\|_{HS}},m,r,d\right)
\nonumber \\ 
& \leq & (t+1) \cdot 
e \! \left( \smfrac{\delta}{\|\Sigma\|_{HS}\sqrt{t}},m',r,d\right)  
\eea
\eel
where the probability is taken over measurement outcomes. 

{\it Proof .-} Here triangle inequality and union bound to
inequalities (\ref{Ineq1}), (\ref{Ineq2}) suffices together with the
properties of $e(\delta,n,r,d)$.  Note that the indirect and direct
measurements are performed on disjoint subsystems, so that there is a
probability space for the joint outcomes.  

% {\bf Remark/Comment: One needs here both decreasing property with
% respect to $\delta$ and $m$. Decreasing in $m$ and this is true but
% for $m$ large enough that linear term dominates the logharitmic
% one. Maybe one shold coment on that? P.}.

\subsubsection{Passing from $\Romr$-s to their integrals and then to 
a close-by state}

Note that both integration and the measurement of a state to produce
the classical distribution of the outcomes are both linear, completely
positive, and trace-preserving maps.  
Thus, Lemma \ref{lem:SigmaDistance} still holds under the replacement
$\Romr \rightarrow \int \Romr d\mu(\sigma)$.
Furthermore, if 
\be
\left \| \varrho_{2m} - \int \Romr d \mu(\sigma) \right \|_{\rm tr} 
\leq \epsilon \,.
\ee
We can use the fact that the trace distance is nonincreasing under the
measurement (a TCP map) to prove the following: 
 \bel
 \label{lem:StateDistance}
 For a state $\varrho_{2m}$ of $2m$ systems 
 satisfying $\| \varrho_{2m} - \int \Romr d \mu (\sigma) \|_{\rm tr}
 \leq \epsilon $
 we have 
 \bea
 & & \Pr \left( \left|\SgmAemp-\SgmBemp \right|>2\delta \right) 
\nonumber
\\  & \leq & 
 (t+1) \cdot e \left( \smfrac{\delta}{\|\Sigma\|_{HS}\sqrt{t}},m',r,d\right)
 +\epsilon \,.
 \eea
 where the probability is evaluated over the probability distribution
 ${\cal P}'$ on outcomes of measurement ${\cal L}_{1}^{\otimes m'}
 \otimes ...\otimes {\cal L}_{t}^{\otimes m'} \otimes {\cal
 M}^{\otimes m}$ induced by the state $\varrho_{2m}$.
\eel

%%%%%%%%%%%%%%%%%%%%%%%%%%%%%%%%%%%%%%%%%%%%%%%%%%%%%%%%
\subsubsection{Inferring direct average 
on $n+m$ samples of general state $\varrho_{2m+n}$ from 
indirect measurements on $m$ samples}
%%%%%%%%%%%%%%%%%%%%%%%%%%%%%%%%%%%%%%%%%%%%%%%%%%%%%%%%
Now we pass to the general permutationally invariant 
state $\varrho_{2m+n}$.  We have the following: 
\bet
Consider permutationally invariant state $\varrho_{2m+n}$ on
$\hcal^{\ot 2m+n}$ and $\dim \hcal = d$. On this state we perform the
measurement ${\cal L}_{1}^{\otimes m'} \otimes \cdots \otimes {\cal
L}_{t}^{\otimes m'} \otimes {\cal M}^{\otimes m+n}$ which induces the
probability measure ${\cal P}''$. (Note that $\pcal'$ is simply the
marginal of $\pcal''$.)  Evaluating the probability over $\pcal''$, 
we have
\be
 \Pr(|\SgmBemp-\SgmAAemp)|> 3 \delta) \leq e_{1} + e_{2} +e_{3}
\ee
where 
\bea 
e_{1} &=&2e^{-{n(r+1)\over 2(2m+n)} + {1\over 2} d^{\bf 2} \ln n } 
\nonumber \\
e_{2}&=&(t+1)
%e(\frac{\delta}{||\Sigma||_{HS}\sqrt{t}},k,r,d),
2^{-(\frac{\delta^{2}}{4t||\Sigma||_{HS}^{2}}-H(\frac{r}{m'})) m'  
+ d \log(\frac{m'}{2}+1)} ~~~{\rm and} 
\nonumber \\
e_{3}&=&d e^{-\frac{m\delta^{2}}{ {\bf 8}d||\Sigma||_{HS}^{2}}}.
\eea
\label{th:box2}
\eet
{\it Proof -} The parameters $e_{1}$, $e_{2}$, $e_{3}$ come 
from the generalized quantum de Finetti theorem, the Chernoff bound 
and the sampling proposition respectively. 

To start, we apply item $(1)$ of Theorem \ref{thm:fincher} to 
$\varrho_{2m}={\rm Tr}_{n}\varrho_{2m+n}$ to obtain $\|\varrho_{2m}-\int
\Romr d \mu(\sigma)\|_{\rm tr} \leq \epsilon $ with $\epsilon=e_{1}$.
Then, we apply Lemma \ref{lem:StateDistance} to get 
\be
 \Pr \, (|\SgmAemp-\SgmBemp|>2 \delta) \leq e_{1} + e_{2}
\label{e12}
\ee

Now we need to connect $\SgmBemp$ with $\SgmAAemp$.  For this we need
the fact that $\mcal$ has at most $d$ outcomes, and we need 
the random sampling theorem, Proposition \ref{prop:srodka}, which gives 
$\Pr \, (\|Q_{\Sigma}^{m} - Q_{\Sigma}^{m+n}\|_{\rm tr} > \delta )\leq d
e^{-{m \delta^2/8 d}}$ where $Q_{\Sigma}^{m}$ is the relative
frequency distribution on outputs of $\mcal$ induced by the state $\rho_m$
(partial trace of $\rho_{2m+n}$ over $m+n$ systems and
$Q_{\Sigma}^{m+n}$ is the relative frequency distribution induced on
the outcomes of $\mcal$ by the state $\rho_{m+n}$ (partial trace of
$\rho_{2m+n}$ over $m$ systems) and $d$ is the dimension of the
elementary Hilbert space ${\cal H}$ (thus $\varrho_{2m}$ is defined on
${\cal H}^{\otimes 2m}$.  Using Lemma \ref{closeaverages} (taking $t=1$ here 
for the direct measurement) we go to the
averages \be \Pr \, (|\SgmAemp-\SgmAAemp|>3\delta) \leq e_{3}.
\label{e3}
\ee
Applying the union bound to Eqs.\ (\ref{e12}) and (\ref{e3})
we obtain the statement of the theorem.

% Recall that the parameter $t$ is a constant, since (it is number of
% observables $L_{i}$ to which we have decomposed the observable
% $\Sigma$).

\noindent {\bf Acknowledgments} We thank Daniel Gottesman and
Hoi-Kwong Lo for valuable discussions. KH acknowledges support of the
Foundation for Polish Science. MH is supported by EC IP SCALA
IST-015714.  JO acknowledges support from the Royal Society and EU
grant QAP IST-015848.  DL is supported by the CRC, CRC-CFI, ORF, CIAR,
NSERC, MITACS, and ARO.  Part of this work was initiated during the
QIS programme at the Isaac Newton Institute (2004).
%\end{acknowledgments}

% \bibliographystyle{IEEEsty}
% \bibliography{be-key}

\begin{thebibliography}{10}

\bibitem{HHHO03}
Karol Horodecki, Micha\l{} Horodecki, Pawe\l{} Horodecki, and Jonathan
  Oppenheim,
\newblock ``Secure key from bound entanglement,''
\newblock {\em Phys. Rev. Lett.}, vol. 94, pp. 160502, 2005.

\bibitem{EPR}
A.~Einstein, B.~Podolsky, and N.~Rosen,
\newblock ``Can quantum-mechanical description of physical reality be
  considered complete?,''
\newblock {\em Phys. Rev.}, vol. 47, pp. 777, 1935.

\bibitem{SP00}
P.~Shor and J.~Preskill,
\newblock ``Simple proof of security of the bb84 quantum key distribution
  protocol,''
\newblock {\em Phys. Rev. Lett.}, vol. 85, pp. 441--444, 2000,
\newblock quant-ph/0003004.

\bibitem{LC99}
H.-K. Lo and H.~F. Chau,
\newblock ``Unconditional security of quantum key distribution over arbitrarily
  long distances,''
\newblock {\em Science}, vol. 283, pp. 2050--2056, 1999,
\newblock quant-ph/9803006.

\bibitem{LCA01}
H.-K. Lo, H.~Chau, and M.~Ardehali,
\newblock ``Efficient quantum key distribution scheme and proof of its
  unconditional security,''
\newblock {\em J. of Cryptology}, vol. 18, pp. 133--165, 2005.

\bibitem{Renner05}
R.~Renner,
\newblock {\em Security of Quantum Key Distribution},
\newblock Ph.D. thesis, ETH, Zurich, 2005.

\bibitem{Bennett94a}
C.~Bennett, G.~Brassard, R.~Jozsa, D.~Mayers, A.~Peres, B.~Schumacher, and
  W.~Wootters,
\newblock ``Reduction of quantum entropy by reversible extraction of classical
  information,''
\newblock {\em Journal of Modern Optics}, vol. 41(12), pp. 2307--2314, 1994.

\bibitem{BHLMO04}
Michael Ben-Or, Michal Horodecki, Debbie~W. Leung, Dominic Mayers, and Jonathan
  Oppenheim,
\newblock ``The universal composable security of quantum key distribution.,''
\newblock in {\em Theory of Cryptography: Second Theory of Cryptography
  Conference, TCC 2005}, Joe Kilian, Ed. 2005, vol. 3378 of {\em Lecture Notes
  in Computer Science}, pp. 386--406, Springer-Verlag.

\bibitem{BM02}
M.~Ben-Or and D.~Mayers,
\newblock ``Composing quantum and classical protocols,''
\newblock quant-ph/0409062.

\bibitem{BB84}
C.~Bennett and G.~Brassard,
\newblock ``Quantum cryptography: {P}ublic key distribution and coin tossing,''
\newblock in {\em Proceedings of IEEE International Conference on Computers,
  Systems and Signal Processing}, New York, 1984, pp. 175--179, IEEE,
\newblock Bangalore, India, December 1984.

\bibitem{Bruss98}
D.~Bruss,
\newblock ``Optimal eavesdropping in quantum cryptography with six states,''
\newblock {\em Phys. Rev. Lett.}, vol. 81, pp. 3018--3021, 1998.

\bibitem{DEJMPS96}
D.~Deutsch, A.~Ekert, R.~Jozsa, C.~Macchiavello, S.~Popescu, and A.~Sanpera,
\newblock ``Quantum privacy amplification and the security of quantum
  cryptography over noisy channels,''
\newblock {\em Phys. Rev. Lett.}, vol. 77, pp. 2818, 1996,
\newblock quant-ph/9604039.

\bibitem{BBCM95}
C.~Bennett, G.~Brassard, C.~Cr\'epeau, and U.~Maurer,
\newblock ``Generalized privacy amplification,''
\newblock {\em IEEE Trans.\ Inf.\ Th.}, vol. 41, pp. 1915--1923, 1995.

\bibitem{Mayers96}
D.~Mayers,
\newblock ``Quantum key distribution and string oblivious transfer in noisy
  channels,''
\newblock in {\em Advances in Cryptography--Proceedings of Crypto'96}, New
  York, 1996, pp. 343--357, Springer-Verlag.

\bibitem{E91}
A.~Ekert,
\newblock ``Quantum cryptography based on {B}ell's theorem,''
\newblock {\em Phys. Rev. Lett.}, vol. 67(6), pp. 661--663, 1991.

\bibitem{BDSW96}
Charles~H. Bennett, David~P. DiVincenzo, John Smolin, and William~K. Wootters,
\newblock ``Mixed-state entanglement and quantum error correction,''
\newblock {\em Phys. Rev. A}, vol. 54, pp. 3824--3851, 1997.

\bibitem{huge-key}
Karol Horodecki, Michal Horodecki, Pawel Horodecki, and Jonathan Oppenheim,
\newblock ``General paradigm for distilling classical key from quantum
  states,''
\newblock quant-ph/0506189.

\bibitem{DW03}
Igor Devetak and Andreas Winter,
\newblock ``Distillation of secret key and entanglement from quantum states,''
\newblock {\em Proc. R. Soc. Lond. A}, 2005.

\bibitem{Peres96}
Asher Peres,
\newblock ``Separability criterion for density matrices,''
\newblock {\em Phys. Rev. Lett.}, vol. 77, pp. 1413, 1996.

\bibitem{9801069}
M.~Horodecki, P.~Horodecki, and R.~Horodecki,
\newblock ``Mixed-state entanglement and distillation, is there a ``bound''
  entanglement in nature?,''
\newblock {\em Phys. Rev. Lett.}, vol. 80, pp. 5239--5242, 1998.

\bibitem{HLLO05}
Karol Horodecki, Debbie Leung, Hoi-Kwong Lo, and Jonathan Oppenheim,
\newblock ``Quantum key distribution based on arbitrarily-weak distillable
  entangled states,''
\newblock {\em Phys. Rev. Lett.}, 2006.

\bibitem{GL01}
G.~Gottesman and H.-K. Lo,
\newblock ``Proof of security of quantum key distribution with two-way
  classical communications,''
\newblock {\em IEEE Transactions on Information Theory}, vol. 49, no. 2, pp.
  457--475, 2003,
\newblock quant-ph/0105121.

\bibitem{RK04b}
R.~Renner and R.~Koenig,
\newblock ``A de finetti representation for finite symmetric quantum states,''
\newblock {\em J. Math. Phys.}, vol. 46, 2005,
\newblock quant-ph/0410229.

\bibitem{JC}
J.~C. Boileau,
\newblock Personal communication, September 2007.

\bibitem{AB02}
H.~Aschauer and H.J. Briegel,
\newblock ``(1) secure quantum communication over arbitrary distances or (2)
  entanglement purification with noisy apparatus can be used to factor out an
  eavesdropper?,''
\newblock {\em Phys. Rev. A}, vol. 66, pp. 032302, 2002,
\newblock quant-ph/(1) 0008051 or (2) 0108060.

\bibitem{RGK05}
R.~Renner, N.~Gisin, and B.~Kraus,
\newblock ``An information-theoretic security proof for QKD protocol,'' 
\newblock {\em Phys. Rev. A}, vol. 72, pp. 012332, 2005.

\bibitem{PL05}
J.~Preskill and H.-K. Lo,
\newblock ``Phase randomization improves the security of quantum key
  distribution,'' 2005,
\newblock quant-ph/0504209.

\bibitem{RSS06}
G.~Smith, J.~Renes, and J.~Smolin,
\newblock ``Better codes for bb84 with one-way post-processing,''
\newblock quant-ph/0607018.

\bibitem{CRE04}
M.~Christandl, R.~Renner, and A.~Ekert,
\newblock ``A generic security proof for quantum key distribution,''
\newblock quant-ph/0402131.

\bibitem{QIP06}
K.~Horodecki, M.~Horodecki, P.~Horodecki, D.~W. Leung, H.-K. Lo, and
  J.~Oppenheim,
\newblock ``Unconditionally secure privacy using channels that cannot convey
  quantum information,'' Presented by K. Horodecki, QIP 2006, Paris.

\bibitem{RS06}
J.~Renes and G.~Smith,
\newblock ``Noisy preprocessing and the distillation of twisted states,''
\newblock {\em Phys. Rev. Lett.}, 2007,
\newblock quant-ph/0603262.

\bibitem{KGR05}
B.~Kraus, N.~Gisin, and R.~Renner,
\newblock ``Lower and upper bounds on the secret key rate for \mbox{QKD}
  protocols using one--way classical communication,''
\newblock {\em Phys. Rev. Lett.}, vol. 95, pp. 080501, 2005.

\bibitem{bigkey-shortpaper}
K. Horodecki, M.~Horodecki, P.~Horodecki, D.~Leung, and J.~Oppenheim,
\newblock ``Unconditional privacy over channels which cannot convey quantum
  information,''
\newblock to appear in Phys. Rev. Lett, quant-ph/0702077.

\bibitem{derive-lca01} Using page 34 of \cite{LCA01} (and realizing
  that $A(\lambda,p) \approx (p{-}\lambda)^2/(p\,(1{-}p)\ln 2)$ to
  lowest order in $p{-}\lambda$ (note that $\lambda \leq p$ is not
  necessary) and running the argument for both black and white balls,
  we get a $2$-sided bound. Let $m$ be the sample size, $n$ be the
  total number of systems, and $\e_{xP},\e_{zP}$, $\e_{x},\e_{z}$ as
  defined in the text. Then, \be {\rm Pr}( |\e_{xP}{-}\e_{x}| \geq
  \delta) \leq 2 \cdot \exp \lpm -m (4 \delta^2 - \smfrac{m}{n-m})
  \rpm . \ee If $m < (\smfrac{2\delta^2}{1{+}2\delta^2}) n$ (usually
  $m \sim \sqrt{n}$ or $\log n$), then, \be {\rm Pr}(\,
  |\e_{xP}{-}\e_{x}| \geq \delta) \leq 2 \cdot \exp \lpm -2 m \delta^2
  \rpm . \ee

\bibitem{Uhlmann76}
A.~Uhlmann,
\newblock ``The `transition probability' in the state space of a
  $\,^*$-algebra,''
\newblock {\em Reports on Mathematical Physics}, vol. 9, pp. 273--279, 1976.

\bibitem{Hamada03}
M.~Hamada,
\newblock ``Reliability of calderbank-shor-steane codes and security of quantum
  key distribution,''
\newblock {\em J. of Phys. A: Mathematical and General}, , no. 34, 2004,
\newblock quant-ph/0308029.

\bibitem{HPHH05}
Karol Horodecki, Lukasz Pankowski, Michal Horodecki, and Pawel Horodecki,
\newblock ``Low dimensional bound entanglement with one-way distillable
  cryptographic key,''
\newblock quant-ph/0506203.

\bibitem{HHH99}
Pawel Horodecki, Michal Horodecki, and Ryszard Horodecki,
\newblock ``Binding entanglement channels,''
\newblock {\em Phys. Rev. Lett.}, vol. 47, pp. 347--354, 2000.

\end{thebibliography}

\end{document}